\documentclass{emulateapj}

\usepackage{mathtools}
\usepackage{epsfig}
\usepackage{tabularx}
\usepackage{multirow}
\usepackage{subfigure}
\usepackage{rotating}
\usepackage{booktabs}
\usepackage{array}
\usepackage{xfrac}
\usepackage{natbib}

\newcolumntype{R}{>{\raggedleft\arraybackslash}X}
\newcolumntype{C}{>{\centering\arraybackslash}X}

\def\iso#1#2{\mbox{${}^{#2}{\rm #1}$}}
\def\be1#1{\iso{Be}{1#1}}
\def\fe6#1{\iso{Fe}{6#1}}
\def\pu2#1#2{\iso{Pu}{2#1#2}}

\def\pfrac#1#2{\left( \frac{#1}{#2} \right)}

\def\msol{M_\odot}

\def\RE{R_\oplus}
\def\ucrust{U_{\rm crust}}
\def\usediment{U_{\rm sediment}}

\def\na{New~Astron.} 
\def\nar{New~Astron.~Rev.} 

\def\beq{\begin{equation}}
\def\eeq{\end{equation}}
\def\beqar{\begin{eqnarray}}
\def\eeqar{\end{eqnarray}}

\begin{document}

\title{Radioactive Iron Rain: \\ 
\vskip 0.075in
Transporting \fe60 in Supernova Dust to the Ocean Floor
\vskip 0.15in
}

\author{Brian J. Fry and Brian D. Fields}

\affil{Department of Astronomy, University of Illinois, Urbana, IL 61801, USA}

\author{John R. Ellis}

\affil{Theoretical Physics and Cosmology Group, Department of Physics, King's College London,
London WC2R 2LS, UK; \\  Theory Department, CERN, CH-1211 Geneva 23, Switzerland}

\begin{abstract} 

Several searches have found evidence of \fe60 deposition, presumably from a near-Earth supernova (SN), with concentrations that vary in different locations on Earth.  This paper examines various influences on the path of interstellar dust carrying \fe60 from a SN through the Heliosphere, with the aim of estimating the final global distribution on the ocean floor.  We study the influences of magnetic fields, angle of arrival, wind and ocean cycling of SN material on the concentrations at different locations.  We find that the passage of SN material through the mesosphere/lower thermosphere (MLT) is the greatest influence on the final global distribution, with ocean cycling causing lesser alteration as the SN material sinks to the ocean floor.  SN distance estimates in previous works that assumed a uniform distribution are a good approximation.  Including the effects on surface distributions, we estimate a distance of $46^{+10}_{-6}$ pc for a $8-10 \ \msol$ SN progenitor.  This is consistent with a SN occurring within the Tuc-Hor stellar group $\sim$2.8 Myr ago with SN material arriving on Earth $\sim$2.2 Myr ago.  We note that the SN dust retains directional information to within $1^{\circ}$ through its arrival in the inner Solar System, so that SN debris deposition on inert bodies such as the Moon will be anisotropic, and thus could in principle be used to infer directional information.  In particular, we predict that existing lunar samples should show measurable \fe60 differences.

\begin{center}
{\tt KCL-PH-TH/2016-15}, {\tt LCTS/2016-09}, {\tt CERN-TH-2016-076}
\end{center}

\end{abstract}

\section{Introduction}
\label{sect:intro}

Supernovae (SNe) are some of the most spectacular 
explosions in our Galaxy.  
Occurring at a rate of $\sim$$1-3$ per century in the Milky Way \citep[e.g.,][and references 
therein]{adams2013}, it is likely that one (if not more) has exploded close enough to have 
produced detectable effects on the Earth.
Speculation on biological effects of a near-Earth SN has a long history in the literature
\citep[e.g.,][]{shkl68,alva80,es95}, and \citet{efs96} and \citet{kor96} proposed using radioactive isotopes such as \fe60 and \pu244 to
find direct evidence of such an event.  Although several studies have searched for \pu244,
this paper will focus exclusively on \fe60.  For more recent examinations of 
\pu244, see \citet{wall00,wall04} and \citet{wall15a}.

With this motivation, \citet{knie99} examined a sample of ferro-manganese (Fe-Mn) 
crust from Mona Pihoa in the South Pacific and found an anomaly in \fe60 concentration that 
suggested a SN occurred near Earth sometime within the last 5 Myr (a specific time could not be determined).  
The study was later expanded in \citet{knie04} using a different Fe-Mn crust sample from the equatorial Pacific Ocean floor, 
and found a distinct signal in \fe60 abundance $\sim2.2$ Myr ago, with a \fe60 fluence, ${\cal F}$, at the time of arrival
calculated to have been  ${\cal F}_{\rm Knie}=1.41 \times 10^{6} \ {\rm atoms \ cm^{-2}}$.  \citet{fit08} subsequently
confirmed the detection by Knie in the Fe-Mn crust, but did not find a corroborating signal in sea sediment 
samples from the northern Atlantic Ocean.  Fitoussi et al. noted several reasons for the discrepancy,
including variations in the background and differences in the uptake efficiencies between the Fe-Mn crust and
sediment.  An excess of \fe60 has also been found in lunar regolith samples \citep{cook09,fimi12,fimi14,fimi16}
but, due to the nature of the regolith, only the presence of a signal is detectable, not the precise 
arrival time or fluence \citep{feige13}.  Subsequently, results from Eltanin sediment samples from the southern Indian Ocean
were reported in \citet{feige14}, confirming the \citet{knie04} Fe-Mn crust detection in these sea sediment samples
and leading to an estimated arrival fluence of ${\cal F}_{\rm Feige} = 1.42 \times 10^{7} \ {\rm atoms \ cm^{-2}}$.
\footnote{It should be noted this is the fluence for the period that overlaps the \citet{knie04} detection.  
\citet{feige14} found the signal to extend in time beyond the \citet{knie04} time interval with a total time-integrated 
fluence of ${\cal F}_{\rm Feige} = (2.32 \pm 0.60) \times 10^{7} \ {\rm atoms \ cm^{-2}}$.  In addition, \citet{wall16} found a larger 
total time-integrated value of ${\cal F}_{\rm Wallner} = (3.5 \pm 0.2) \times 10^{7} \ {\rm atoms \ cm^{-2}}$.  For the purposes of this paper 
we will focus solely on the fluences that overlap with the \citeauthor{knie04} fluence.}
This fluence is an order of magnitude higher than found by \citet{knie04}, and the difference in fluence values
was attributed to differences in uptake efficiencies for sea sediment versus Fe-Mn crust.  \citet{feige14} and 
\citet{feige13} noted that, whilst the sea sediment uptake efficiency is most likely 
$\usediment \approx 100 \%$, other observations \citep[including the recent, extensive study of \fe60 measurements by ][]{wall16} 
suggest the Fe-Mn crust has an uptake efficiency of $\ucrust \in [0.1, 1]$.

Complementing the multiple searches for \fe60 and other isotopes, several papers have discussed the 
interpretations and implications of the \fe60 signal.  The hydrodynamic models used by \citet{faj08} 
discussed the interaction of a SN blast with the solar wind, and highlighted the necessity \citep[see also][]{af11} 
of ejecta condensation into dust grains capable of reaching Earth.  \citet{fry15} examined the possible sources of the 
Knie \fe60 signal, finding an Electron-Capture SN (ECSN), with Zero-Age Main Sequence (ZAMS) mass 
$\approx 8-10 \ \msol$ (``$\odot$'' refers to the Sun), to be the most likely progenitor, while not completely ruling out a 
Super Asymptotic Giant Branch (SAGB) star with ZAMS mass $\approx 6.5-9 \ \msol$.

With regards to a possible location of the progenitor, \citet{bene2002} suggested that the source event for the \fe60 
occurred in the Sco-Cen OB association.  This association was $\sim$130 pc away at the time of the \fe60-producing event, 
and its members were described in detail by \citet{fuchs06}.  \citet{breit12} modeled the formation of the Local Bubble with 
a moving group of stars (approximating the Sco-Cen association) and plotted their motion in the Milky Way at 
5-Myr intervals for the past 20 Myr \citep[see Figure 9 of][]{breit12}.  More recently, \citet{breit16} have expanded this examination
using hydrodynamic simulations to model SNe occurring within the Sco-Cen association and track the \fe60 dust entrained
within the blast.  Additionally, \citet{kach15} and \citet{sav15} 
found a signature in the proton cosmic ray spectrum suggesting an injection of cosmic rays associated with a 
SN occurring $\sim$2 Myr ago, and  \citet{binns16} found \fe60 cosmic rays, suggesting a SN 
origin within the last $\sim 2.6$ Myr located $\lesssim 1$ kpc of Earth, based on the \fe60 lifetime and cosmic ray diffusion. 
With particular relevance for our discussion, \citet{mam16} suggested the Tuc-Hor  group could have provided an ECSN to 
produce the \fe60.  The group was within $\sim$60 pc of Earth $\sim$2.2 Myr ago and, given the masses of the 
current group members, could well have hosted a star with a ZAMS mass $\geq 8 \msol$.  

\citet{fry15} noted that these and other studies assumed a uniform deposition of \fe60 material over Earth's entire surface, 
and proposed that the direction of arriving material and the Earth's rotation could shield portions of Earth's surface from 
SN material.  Since \fe60 dust from a SN would be arriving along one direction instead of isotropically, 
the suggestion was that certain portions of Earth's surface would face the SN longer than others and collect 
more arriving material.  This could explain why the northern \citet{fit08} sediment samples showed no obvious signal, 
whereas the southern \citet{feige13}, \citet{feige14} sediment samples showed a stronger signal than the equatorial \citet{knie04}
crust sample.  

This paper re-examines that possibility, and studies how the angle of arrival of dust from a SN effects the 
deposition on the Earth's surface.  We show that the dust propagation in the inner Solar System introduces
deflections of order a few degrees.  Thus, the angle of arrival drastically changes the received 
fluence at the top of the Earth's atmosphere.  However,
any such variations are lost as the SN material descends through our
atmosphere, and the final global distribution is due primarily to atmospheric influences with slight alterations due to ocean cycling.
This confirms an isotropic deposition on the Earth's surface as a reasonable assumption when making order of magnitude calculations.
This in turn removes an uncertainty in estimates of the distance to the \fe60 progenitor, which may have been within the Sco-Cen or Tuc-Hor stellar groups.

In contrast, the memory of the angle of arrival would be retained in deposits on airless Solar System
bodies such as the Moon.  We find that lunar samples should show significant variation in
SN \fe60 abundance if the source was in the Tuc-Hor or the Sco-Cen groups.
Thus the \fe60 pattern on the Moon in principle can give directional information, serving as a
low-resolution ``antenna'' that could potentially test proposed source directions.

Lastly, our examination assumes the passage of a single SN.  In studying Solar System/terrestrial influences on 
SN \fe60, we find that none are capable of extending the signal postulated by \citet{fry15} to the wider signal detected by \citet{feige14} 
and \citet{wall16}.  This supports the assertion by \citet{breit16} and \citet{wall16} of multiple SNe producing the \fe60 signal.

\section{Motivation}
\label{sect:motivation}

\citet{fry15} defined the decay-corrected fluence as that measured at the time the signal arrived.\footnote{Other 
descriptions of fluence have been used in the literature, but here we deal exclusively with the 
arrival/decay-corrected fluence.  For a full description see \citet{fry15}.}  However, inherent in the formula used in
\citet{fry15} (and in all other studies known to us) was the assumption that the material was distributed
uniformly, that is, {\em isotropically}, over Earth's entire surface.
Here we examine this assumption in detail.  In fact, the arriving SN blast will be highly directional,
roughly a plane wave on Solar System scales \citep{faj08}.

In this paper, we will assume that all SN dust will be entrained in the blast plasma
as it arrives in the Solar System.  That is, we ignore any relative motion of the dust in the blast.
\footnote{More precisely, we assume that any velocity dispersion among dust particles and relative to the plasma 
will be small compared to the bulk plasma velocity.  We will relax this assumption in a forthcoming paper.}
Thus the dust will arrive with the same velocity vector as the blast.
The SN dust particles will then encounter the blast/solar wind interface, decouple, and be injected
into the Solar System with a plane-wave geometry.  

As SN dust traverses the Solar System, it passes through magnetic fields, multiple layers of the Earth's atmosphere
and water currents until finally being deposited on the ocean floor.  In addition, because we would expect dust from a SN to 
arrive as a plane wave as the Earth rotates, different regions would have become exposed to the wave for different durations.  Relaxing the assumption of uniformly distributed debris deposition gives:\footnote{The subscript $i$ sometimes appears in the literature \citep[see e.g.,][]{fry15}.  This refers to the specific isotope/element being examined, but for this paper, we will be examining \fe60 only, so the subscript is not used here.}
\begin{align}
{\cal F}({\it lat, lon}) &= \psi({\it lat, lon}) \nonumber \\ &\times \pfrac{1}{4} \pfrac{M_{\rm ej}}{4 \pi D^{2} A m_{u}} \ U f e^{-t_{\rm travel}/\tau} \, ,
\label{eq:fluence}
\end{align}
where ${\cal F}({\it lat, lon})$ is the fluence of the isotope at the time the signal arrives at a location with latitude and longitude
$({\it lat,lon})$ on Earth's surface.  Here $M_{\rm ej}$ is the mass of the 
ejected isotope, $D$ is the distance the isotope travels from the SN to Earth, $A$ is the atomic mass of the isotope, 
$m_{u}$ is the atomic mass unit, $U$ is the uptake efficiency of the material the isotope is sampled from, 
$f$ is the fraction of the isotope in the form of dust that reaches Earth, $t_{\rm travel}$ is the time taken by the 
isotope to travel from the SN to Earth, and $\tau$ is the mean lifetime of the isotope.  The factor of $\sfrac{1}{4}$ comes from the ratio of
Earth's cross-section to its surface area, and the factor $4\pi$ assumes spherical symmetry in the SN's expansion.  The uptake efficiency is a measure of how readily a material incorporates the elements 
deposited on it.  Sediment accepts nearly all deposited elements, so we assume $\usediment = 1$.
However, the Fe-Mn crust incorporates iron through a chemical leaching process, so the uptake for iron into the 
crust is thought to lie in the range $\ucrust \approx 0.1-1$ \citep[for more discussion, see][]{feige12, feige14, fry15, wall16}.  
In order to account for concentrations and dilutions in the deposition of SN material, we include a factor $\psi$ to
represent the deviation from a uniform distribution $(\psi = 1)$, where $\psi \in [0, \ 1)$ implies a 
diluted deposition and $\psi > 1$ implies a concentrated deposition.

When we compare samples from different terrestrial locations, most of the quantities in Equation (\ref{eq:fluence}) disappear,
so that the fluence ratios depend only on the uptake and distribution factors:
\begin{align}
\label{eq:obsfluence}
\frac{{\cal F}_{\rm Fitoussi}}{{\cal F}_{\rm Knie}} &= \frac{\pfrac{\psi_{\rm Fitoussi}}{4} \pfrac{M_{\rm ej}}{4 \pi D^{2} A m_{u}} U_{\rm Fitoussi} f e^{-t_{\rm travel}/\tau}}{\pfrac{\psi_{\rm Knie}}{4} \pfrac{M_{\rm ej}}{4 \pi D^{2} A m_{u}} U_{\rm Knie} f e^{-t_{\rm travel}/\tau}} \nonumber \\
&= \frac{U_{\rm Fitoussi} \psi_{\rm Fitoussi}}{U_{\rm Knie} \psi_{\rm Knie}} \, .
\end{align}
Similarly:
\beq
\frac{{\cal F}_{\rm Fitoussi}}{{\cal F}_{\rm Feige}} = \frac{U_{\rm Fitoussi} \psi_{\rm Fitoussi}}{U_{\rm Feige} \psi_{\rm Feige}} \, ,
\label{eq:fitoussifeige}
\eeq
\beq
\frac{{\cal F}_{\rm Feige}}{{\cal F}_{\rm Knie}} = \frac{U_{\rm Feige} \psi_{\rm Feige}}{U_{\rm Knie} \psi_{\rm Knie}} \, .
\label{eq:kniefeige}
\eeq
Using these relations, we can test a distribution model against observations.

\section{\fe60 Fluence Observations}
\label{sect:observations}

We examine three studies of \fe60 measurements:  \citet{knie04}, \citet{fit08}, and \citet{feige14}.  These studies have considerable overlap in their time periods and greatly varying locations on the Earth.  We do not examine the \citet{wall16} measurements in detail, first, because the bulk of the analysis for this paper was completed and submitted for review prior to the publication of \citet{wall16}, and second, because many of the samples included in \citet{wall16} are either already included in the other studies, do not cover the period around the 2.2-Myr signal, or were drawn from similar latitudes as the other samples.

{\renewcommand{\arraystretch}{1.5}
\begin{table}[t]
\begin{center}
\caption{Model Cases, Uptakes, and Fluences}
\label{tab:cases}
	\begin{tabularx}{0.48\textwidth}{ C C C } \hline \hline
			Case 							& $\ucrust$ 		& $\usediment$ 	\\
			High Uptake 					& 1 				& 1 			\\ 
			Medium Uptake					& 0.5 				& 1 			\\
			Low Uptake						& 0.1				& 1 			\\ \hline
			${\cal F}_{\rm Knie}$ 			& ${\cal F}_{\rm Fitoussi}$ 	& ${\cal F}_{\rm Feige}$			\\
			$(1.41 \pm 0.49) \times 10^{6}$ & $\leq 1.1 \times 10^{8}$ 		& $(1.42 \pm 0.37) \times 10^{7}$	\\ \hline
	\end{tabularx}
	{\raggedright Fluences are given in atoms cm$^{-2}$}
\end{center}
\end{table}}

\subsection{\citet{knie04} Sample}
\label{subsect:KnSamp}

The \citet{knie04} study used the hydrogenous deep-ocean Fe-Mn crust 237KD from 9$^{\circ}$18' N, 
146$^{\circ}$03' W ($\sim$1,600 km/1,000 mi SE of Hawaii).  The crust growth rate is estimated at 2.37 mm Myr$^{-1}$ \citep{fit08}, 
and samples were taken at separations corresponding to 440- and 880-kyr time intervals.  
\citeauthor{knie04} originally estimated that the \fe60 signal occurred
2.8 Myr ago with a decay-corrected fluence of $(2.9 \pm 1.0) \times 10^{6} \ {\rm atoms \ cm^{-2}}$.  However, 
at the time of their analysis, the half-life of \fe60 was estimated to be 1.49 Myr, and the half-life of \be10 (which was used to date 
individual layers) was estimated to be 1.51 Myr.  Current best estimates for these values are
$\tau_{1/2 \text{, \fe60}} = 2.60$ Myr \citep{rug09, wall15b} and $\tau_{1/2 \text{, \be10}} = 1.387$ Myr \citep{chm10, kor10}.  
This changes the estimated signal arrival time to 2.2 Myr ago, and gives a decay-corrected fluence of 
${\cal F}_{\rm Knie} = (1.41 \pm 0.49) \times 10^{6} \ {\rm atoms \ cm^{-2}}$.  Additionally, \citeauthor{knie04} used an iron uptake 
efficiency of $\ucrust = 0.006$, whereas more recent studies suggest that the uptake for the crust is much higher, 
$\ucrust \approx 0.1-1$ \citep{be11, feige14, wall16}.  In this paper, we consider a ``Medium'' case that uses the Knie fluence 
of ${\cal F}_{\rm Knie} = (1.41 \pm 0.49) \times 10^{6} \ {\rm atoms \ cm^{-2}}$ and an uptake of $\ucrust = 0.5$, 
but we also examine the possibilities that the uptake is higher ($\ucrust = 1$) and lower ($\ucrust = 0.1$).  Of special note, 
\citet{feige14} and \citet{wall16} found $\ucrust \in [0.07, 0.17]$; both studies assumed an isotropic terrestrial distribution and found $\ucrust \approx 0.1$ 
by comparing crust and sediment fluences.  Because the sediment samples came from the Indian Ocean, and the crust samples came from the Pacific and Atlantic Oceans, the distribution factor could potentially be pertinent, so we consider a range of $\ucrust$ values.

\subsection{\citet{fit08} Samples}
\label{subsect:FiSamp}

\citet{fit08} performed measurements on both Fe-Mn crust and sea sediment.  The Fitoussi crust sample came from the 
same Fe-Mn crust used by \citet{knie04}, but from a different section of it.  The Fitoussi sea sediment samples are from 
66$^{\circ}$56.5' N, 6$^{\circ}$27.0' W in the North Atlantic ($\sim$400 km/250 mi NE of Iceland).  The average 
sedimentation rate for the samples is 3 cm kyr$^{-1}$, and slices were made corresponding to time intervals of $10-15$ kyr.  
The sediment samples had a density 1.6 g cm$^{-3}$ and an average iron weight fraction 0.5 wt\%.  
\citet{fit08} examined the period $1.68-3.2$ Myr ago, but found no significant \fe60 signal above the background level 
like that found in the Knie crust sample \citep[Figure 3,][]{fit08}.  In an effort to further analyze their results, 
they calculated the running means for the samples using data intervals of 
$\sim$400 and 800 kyr \citep[Figure 4,][]{fit08}.  This allowed the narrower sediment time intervals to be compared 
to the longer crust time intervals.  They also considered the lowest observed sample measurement as the 
background level, rather than the total mean value used initially.  In this instance, they found a signal of 
marginal significance in the 400-kyr running mean centered at $\sim$2.4 Myr of \fe60/Fe$ = (2.6 \pm 0.8) \times 10^{-16}$.  

For this paper, we consider as part of our ``Medium'' scenario a non-detection by \citet{fit08} (in other words, 
${\cal F}_{\rm Fitoussi} = 0 \ {\rm atoms \ cm^{-2}}$).  In addition, we assume an upper limit set by
non-detection of a signal in the \citet{fit08} sediments because of a high sedimentation rate.  
This is motivated by initial \citet{fit08} measurements that found a slightly elevated \fe60 abundance 
at $\sim$2.25 Myr ago, but were not significant because they were not sufficiently above the background \citep{fit08}.  
To determine this upper limit, we first calculate the number density of iron in the sediment \citep{feige12}:
\beq
n_{\rm Fe} = \frac{w N_{A} \rho}{A} \, ,
\eeq
where $w=0.005$ is the weight fraction of iron in the samples, $N_{A}$ 
is Avogadro's number, $\rho = 1.6 \ {\rm g \ cm^{-3}}$ is the mass density of the sample, and 
$A = 55.845 \ {\rm g \ mol^{-1}}$ is the molar mass for iron.  This yields a number density of 
$n_{\rm Fe} = 8.6 \times 10^{19} \ {\rm atoms \ cm^{-3}}$.  Using the marginally significant signal to calculate 
the \fe60 number density, we find  $n_{\text{\fe60}} = 8.6 \times 10^{19} \ {\rm atoms \ cm^{-3}} \cdot 
2.6 \times 10^{-16} = 2.2 \times 10^{4} \ {\rm atoms \ cm^{-3}}$.  An 870-kyr time interval (in order to compare to the
fluence quoted by \citet{knie04} corresponds to a length of 2610 cm, and gives an upper limit on the fluence of 
$5.9 \times 10^{7} \ {\rm atoms \ cm^{-2}}$.  Correcting for radioactive decay gives the following
upper limit on the fluence at the time the signal arrived:
\begin{align}
&{\cal F}_{\rm Fitoussi} \leq \frac{5.9 \times 10^{7} \ {\rm atoms \ cm^{-2}}}{2^{-2.2 \ {\rm Myr}/2.60 \ {\rm Myr}}} \nonumber \\ 
\Rightarrow \qquad &{\cal F}_{\rm Fitoussi} \leq 1.1 \times 10^{8} \ {\rm atoms \ cm^{-2}} \, .
\end{align}

\subsection{\citet{feige14} Samples}
\label{subsect:FeSamp}

\citet{feige14} studied four sea sediment samples from the South Australian Basin in the Indian Ocean 
(1,000 km/620 mi SW of Australia).  Three of the sediment cores cover the time period examined by
\citet{knie04} and \citet{fit08}:  ELT45-21 (39$^{\circ}$00.00' S, 103$^{\circ}$33.00' E), 
ELT49-53 (37$^{\circ}$51.57' S, 100$^{\circ}$01.73' E) and ELT50-02 (39$^{\circ}$57.47' S, 104$^{\circ}$55.69' E).  
They have an average density of 1.35 g cm$^{-3}$, an average iron weight fraction of 0.2 wt\%, 
and sedimentation rates of 4 mm kyr$^{-1}$ for ELT45-21 and ELT50-02 and 3 mm kyr$^{-1}$ for ELT49-53.
\citet{feige14} studied samples from $0-4.5$ Myr ago, primarily in the time period of the Knie signal and was 
able to corroborate it, finding a decay-corrected fluence
${\cal F}_{\rm Feige} = (1.42 \pm 0.37) \times 10^{7} \ {\rm atoms \ cm^{-2}}$.  For our ``Medium'' scenario,
we adopt the \citet{feige14} fluence and assume that the uptake for sediment (for both the
\citet{fit08} and \citet{feige14} samples) is $\usediment = 1$.  In our model comparisons, we
use the location of the ELT49-53 sample.  Table \ref{tab:cases} summarizes the assumptions we use in our modeling.

\section{Deposit Considerations}
\label{sect:model}

{\renewcommand{\arraystretch}{1.5}
\begin{table*}[t]
\begin{center}
\caption{Maximum Trajectory Deflection for Various Grain Parameters}
\label{tab:magdefl}
	\begin{tabularx}{\textwidth}{ c c c C c c c } \hline \hline
			Grain Charge (V)	& Speed (km s$^{-1}$) 	& Grain Radius ($\mu$m)		& $\beta$ 	& IMF Deflection ($^{\circ}$)	& Magnetosphere Deflection ($^{\circ}$)	& Reaches Earth \\
			5 					& 40					& 0.2						& 0.8		& 0.5							& 0.04									& Yes \\
			0.5 				& 40					& 0.2						& 0.8		& 0.3							& 0.005									& Yes \\
			50 					& 40					& 0.2						& 0.8		& 5								& 0.4									& Yes \\
			5 					& 40					& 0.02						& 0.1		& 3								& 5										& Yes \\
			5 					& 40					& 2							& 0.1		& 0.3							& 0.02									& Yes \\
			5 					& 20					& 0.2						& 0.8		& 0.9							& 0.07									& No \\
			5 					& 80					& 0.2						& 0.8		& 0.4							& 0.02									& Yes \\ \hline
	\end{tabularx}
\end{center}
\end{table*}}

As noted above, in this paper we assume that the dust grains are entrained within the SN shock until it reaches the Heliosphere, 
at which time the dust grains decouple from the shock and enter the Solar System, 
where they are affected by the magnetic fields present.  Apart from the Sun's magnetic, gravitational, and radiative influences, we consider only Earth's magnetic and gravitational influences and ignore those of other objects in the Solar System (e.g., the Moon, Jupiter, etc.).
We describe the dust with fiducial values of grain radius $a \geq 0.2 \ \mu$m, charge corresponding to a voltage ${\cal V} = 5$ V, and initial velocity $v_{\rm grain, 0} \geq 40$ km s$^{-1}$.\footnote{These values are based on the findings in \citet{fry15}.  Dust grains are expected to be larger than 0.2 $\mu$m in order to reach Earth, 5 V is a typical voltage for interstellar grains, and 40 km s$^{-1}$ is a typical arrival velocity for the SN shock \citep[Table 3,][]{fry15}}

\subsection{Magnetic Deflection}
\label{subsect:magdefl}

The grains will experience a number of forces upon entering the Solar System:  drag from collisions with the solar wind, radiation pressure from sunlight, gravity from the Sun and Earth, and a Lorentz force from magnetic fields since the grains will most likely be charged.
\citet{af11} studied these effects in detail for SN grains, though with somewhat different SN
parameters than are now favored, primarily due to the possible large revisions in crust uptake values.
Nevertheless, following \citet{af11},
we expect the influence of magnetic fields to be the dominant force for most of the grains traveling through interplanetary space.  With our fiducial SN dust properties, we would not expect drag from the solar wind to affect the dust grains significantly, given that the drag stopping distance $R_{\rm drag}$ is much larger than the size of the Solar System \citep{murr04}:\footnote{This is the stopping distance for a supersonic dust grain.  Although the grains are moving subsonically relative to the Sun, they are supersonic relative to the outward-flowing 
solar wind ($v_{\rm SW} \approx 400 \ {\rm km \ s^{-1}}$).}
\beq
R_{\rm drag} = 1.7 \ {\rm pc} \left( \frac{\rho_{\rm grain}}{3.5 \ {\rm g \ cm^{-3}}} \right) \left( \frac{a}{0.2 \ \mu{\rm m}} \right) \left( \frac{7.5 \ {\rm cm^{-3}}}{n_{\rm H}} \right) \, .
\eeq
The remaining forces (gravitational, radiation, and magnetic) have comparable values.

As noted in \citet{fry15}, for a ratio of the Sun's radiation force ($F_{\rm rad}$) to its gravitational force ($F_{\rm grav}$), $\beta \lesssim 1.3$, 
the dust grains will reach Earth's orbit:
\begin{align}
	\beta \equiv \frac{F_{\rm rad}}{F_{\rm grav}} = 0.8 &\left( \frac{C_{\rm r}}{7.6 \times 10^{-5} \ {\rm g \ cm}^{-2}} \right) \left( \frac{Q_{\rm pr}}{1} \right) \nonumber \\ \times &\left( \frac{3.5 \ {\rm g \ cm}^{-3}}{\rho_{\rm grain}} \right) \left( \frac{0.2 \ {\rm \mu m}}{a} \right) \, ,
\end{align}
where $C_{\rm r}$ is a constant and $Q_{\rm pr}$ is the efficiency of the radiation pressure
on the grain \citep[for more detail see][]{gust94}.

The field strength of the interplanetary magnetic field (IMF, generated by the Sun) varies 
from a value of $B \sim 0.1 \ \mu$G at 100 AU to $B \sim 50 \ \mu$G at 1 AU.  This implies the ratio of the 
magnetic to gravitational force varies over a range that is at least $F_{\rm mag}/F_{\rm grav} \approx 0.03 - 2$:
\begin{align}
	\frac{F_{\rm mag}}{F_{\rm grav}} = 2 &\left( \frac{\cal V}{5 \ {\rm V}} \right) \left( \frac{B}{0.3 \ \mu{\rm G}} \right) \left( \frac{v}{40 \ {\rm km \ s^{-1}}} \right) \nonumber \\ 
	\times &\left( \frac{r}{100 \ {\rm AU}} \right)^{2} \left( \frac{3.5 \ {\rm g \ cm}^{-3}}{\rho_{\rm grain}} \right) \left( \frac{0.2 \ {\rm \mu m}}{a} \right)^{2} \, .
\end{align}

Both the IMF and the Magnetosphere (generated by the Earth) have similar strengths at the surfaces of their respective sources ($B \sim 1$ G), that weaken rapidly further away.  Beyond 1 AU, the IMF is less than 100 $\mu$G, likewise the tail portion of the Magnetosphere asymptotically approaches 100 $\mu$G.  Because the Sun's radiation and gravitational forces are of similar magnitude, but opposite directions, we can estimate the influence of magnetic fields on the incoming SN dust grains before the in-depth numerical discussion below.  If we calculate the gyroradius for our fiducial grain values, we get \citep{murr04}:
\begin{align}
R_{\rm mag} = 28 \ {\rm AU} &\left( \frac{\rho_{\rm grain}}{3.5 \ {\rm g \ cm^{-3}}} \right) \left( \frac{5 \ {\rm V}}{\cal V} \right) \left( \frac{100 \ \mu{\rm G}}{B} \right) \nonumber \\ \times &\left( \frac{v_{\rm grain, 0}}{40 \ {\rm km \ s^{-1}}} \right) \left( \frac{a}{0.2 \ \mu{\rm m}} \right)^{2} \, .
\end{align}

Given the sizes of the Solar System ($\sim$100 AU) and the Magnetosphere ($\sim$1000 $\RE$, ``$\oplus$'' refers to the Earth), we would expect some deflection by the IMF, though not a complete disruption since the IMF weakens by several orders of magnitude beyond 1 AU, whereas the Magnetosphere should cause very little deflection of the dust grains.  The numerical results below confirm this expectation, as summarized in Table \ref{tab:magdefl}.

\subsubsection{Heliosphere Transit}
\label{subsubsect:heliotrans}

The IMF has a shape resembling an Archimedean spiral due to a combination of a frozen-in magnetic field, the Sun's rotation, and an outward flowing solar wind \citep{park63}.  At Earth's orbit, the IMF has a value of $\vec{B}_{r, \theta, \phi} = \left< 30, 0, 30 \right> \ \mu$G \citep{gust94}, with the azimuthal component dominating at larger radii \citep{park58}.  \citet{af11} studied the passage of SN dust grains through the IMF, and calculated their deflection, but for velocities $\geq 100$ km s$^{-1}$.  In this section we expand on \citeauthor{af11}'s treatment by looking at slower initial grain velocities and solving numerically the equations of motion for the dust grain.  
\beq
m_{\rm grain} \frac{d \vec{v}_{\rm grain}}{dt} = \vec{F}_{\rm grav, \odot} + \vec{F}_{\rm rad, \odot} + \vec{F}_{\rm mag, \odot}
\label{eq:eqnmotion}
\eeq
We include the Lorentz force, $\vec{F}_{\rm mag, \odot}$, due to the IMF as well as the Sun's gravity, $\vec{F}_{\rm grav, \odot}$, and radiation, $\vec{F}_{\rm rad, \odot}$, forces.  Grain erosion is not included since the erosion timescale is much longer than the crossing time for the grains; neither are changes in grain charge since we expect the charge remains fairly constant once it enters the Solar System \citep{kim98}.  Our results are in good agreement with the broader and more detailed examination completed by \citet{sterk12, sterk13}

\begin{figure*}[t]
	\centering
	\subfigure[]
		{\includegraphics[width=0.5625\textwidth]{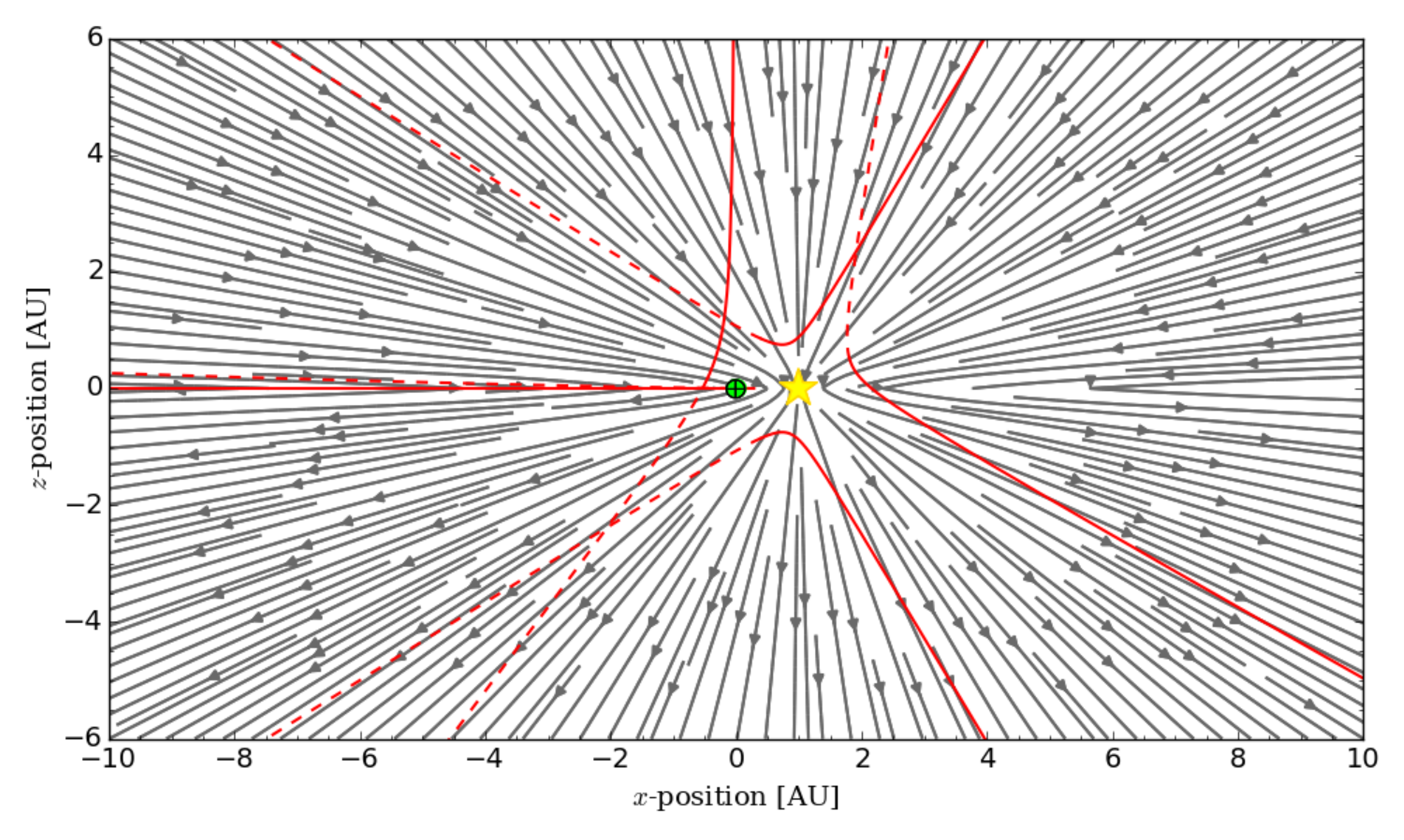}
		\label{fig:HelioSingleXZ}}	
	\subfigure[]
		{\includegraphics[width=0.3375\textwidth]{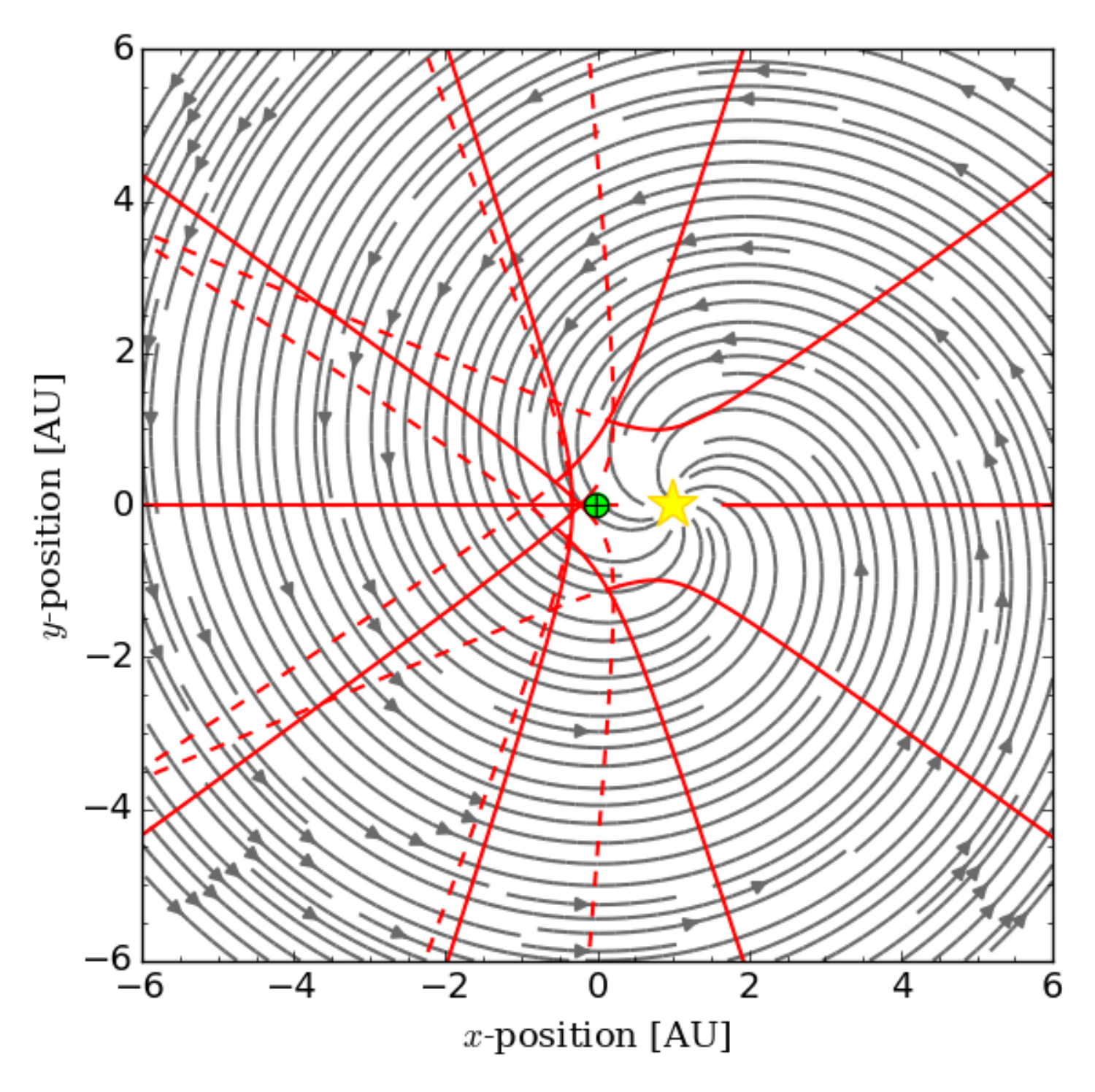}
		\label{fig:HelioSingleXY}} \\
	\subfigure[]
		{\includegraphics[width=0.5625\textwidth]{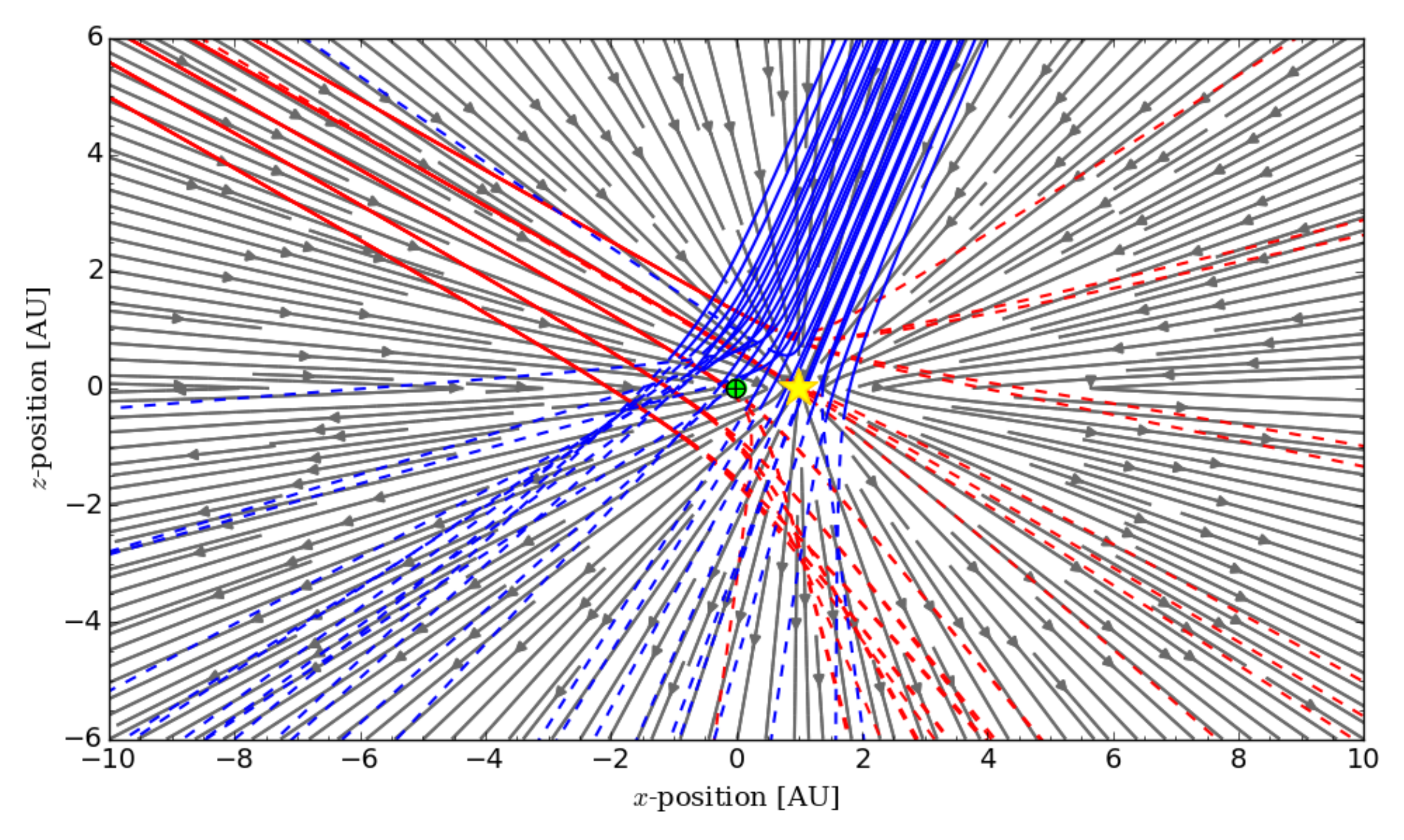}
		\label{fig:HelioMultipleXZ}}	
	\subfigure[]
		{\includegraphics[width=0.3375\textwidth]{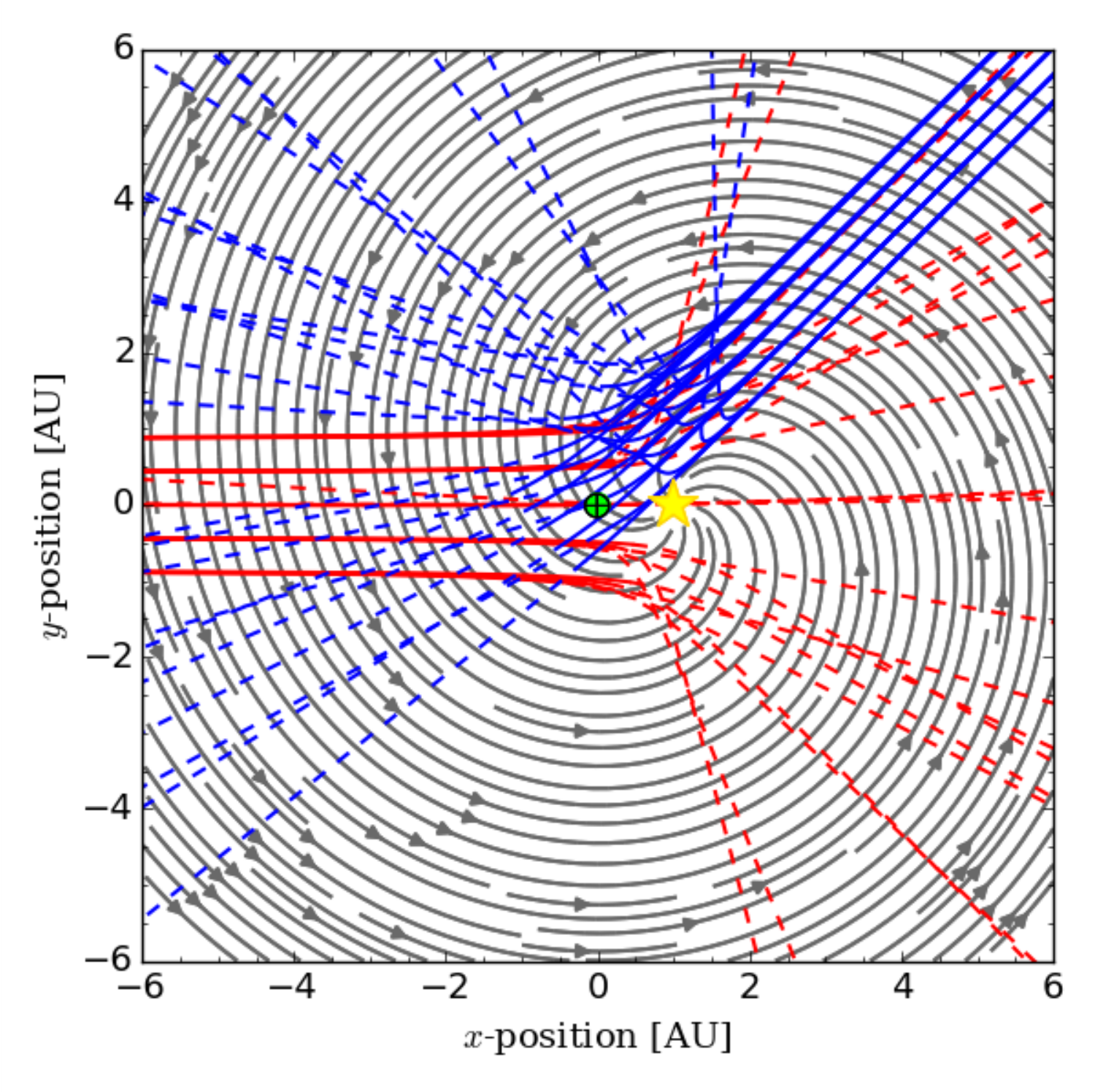}
		\label{fig:HelioMultipleXY}}
	\caption{Sample dust grain trajectories within the Heliosphere.  The magnetic field lines are shown with grey arrows, the Sun is shown by a yellow star at $(1, 0, 0)$ AU, and the Earth is shown by a green $\oplus$ at the origin (NOTE:  the Sun and Earth sizes are not to scale).  Dust grain trajectories are shown in red or blue with the incoming trajectory indicated by solid lines, and, after closest approach to Earth, the outgoing trajectory is indicated by dashed lines.  The initial dust grain parameters are:  $a = 0.2 \ \mu$m, ${\cal V} = 5$ V, and $v = 40$ km s$^{-1}$.  The upper panel shows individual dust grain trajectories initially aimed at Earth and how they behave in the inner Solar System.  The lower panel shows a dust swarm with grains initially travelling parallel to each other, and grains not initially aimed at Earth can be deflected into it.
	\label{fig:HelioTrans}
	}
\end{figure*}

The grains begin 110 AU from the Sun and have initial velocities directed at a location 1 AU away from the Sun representing Earth.  We vary the initial grain directions, speeds, charges, and sizes, and solve for the angle between the grain's initial direction and the line between the grain's starting location and closest approach to Earth's location.  For our fiducial grain values, they experienced $\lesssim 1^{\circ}$ of deflection, and, since their velocities were greater than the solar escape velocity, they continued out of the Solar System after passing Earth's orbit.  Additionally, when examined as a plane wave, the grains showed a fairly uniform deflection amongst neighboring grains until closest approach, meaning that, even though a grain that was initially aimed at Earth would miss by $\sim 1^{\circ}$, another neighboring grain would be deflected into the Earth.  These results suggest that direction information of the grains' source would be retained to within $1^{\circ}$, and that spatial and temporal dilutions/concentrations of the \fe60 signal can be ignored, see Figure \ref{fig:HelioTrans}.

\subsubsection{Earth's Magnetosphere Transit}
\label{subsubsect:magnetotrans}

\begin{figure*}[t]
	\centering
	\subfigure[]
		{\includegraphics[width=0.45\textwidth]{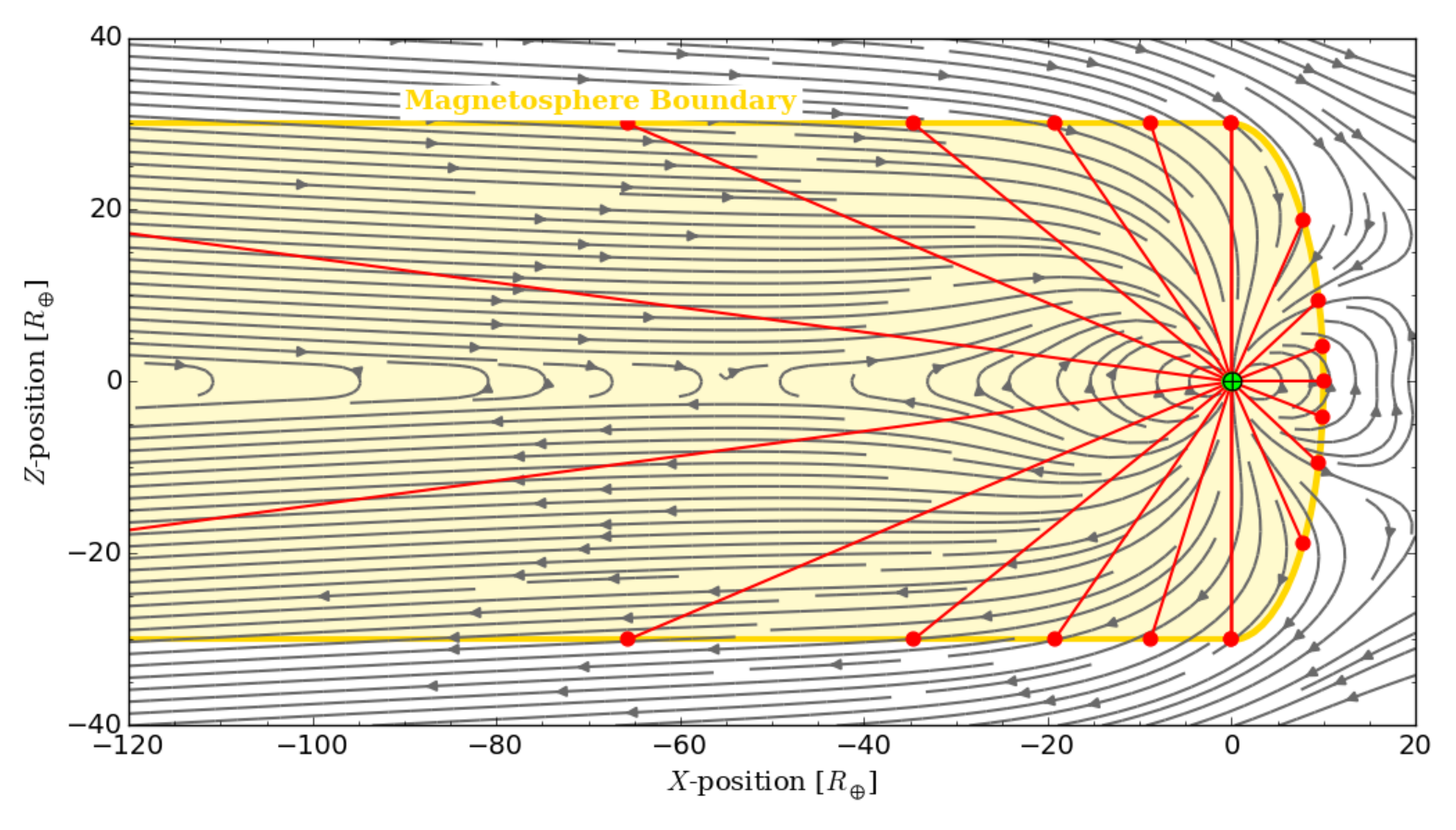}
		\label{fig:MagnetoSingle}} ~
	\subfigure[]
		{\includegraphics[width=0.45\textwidth]{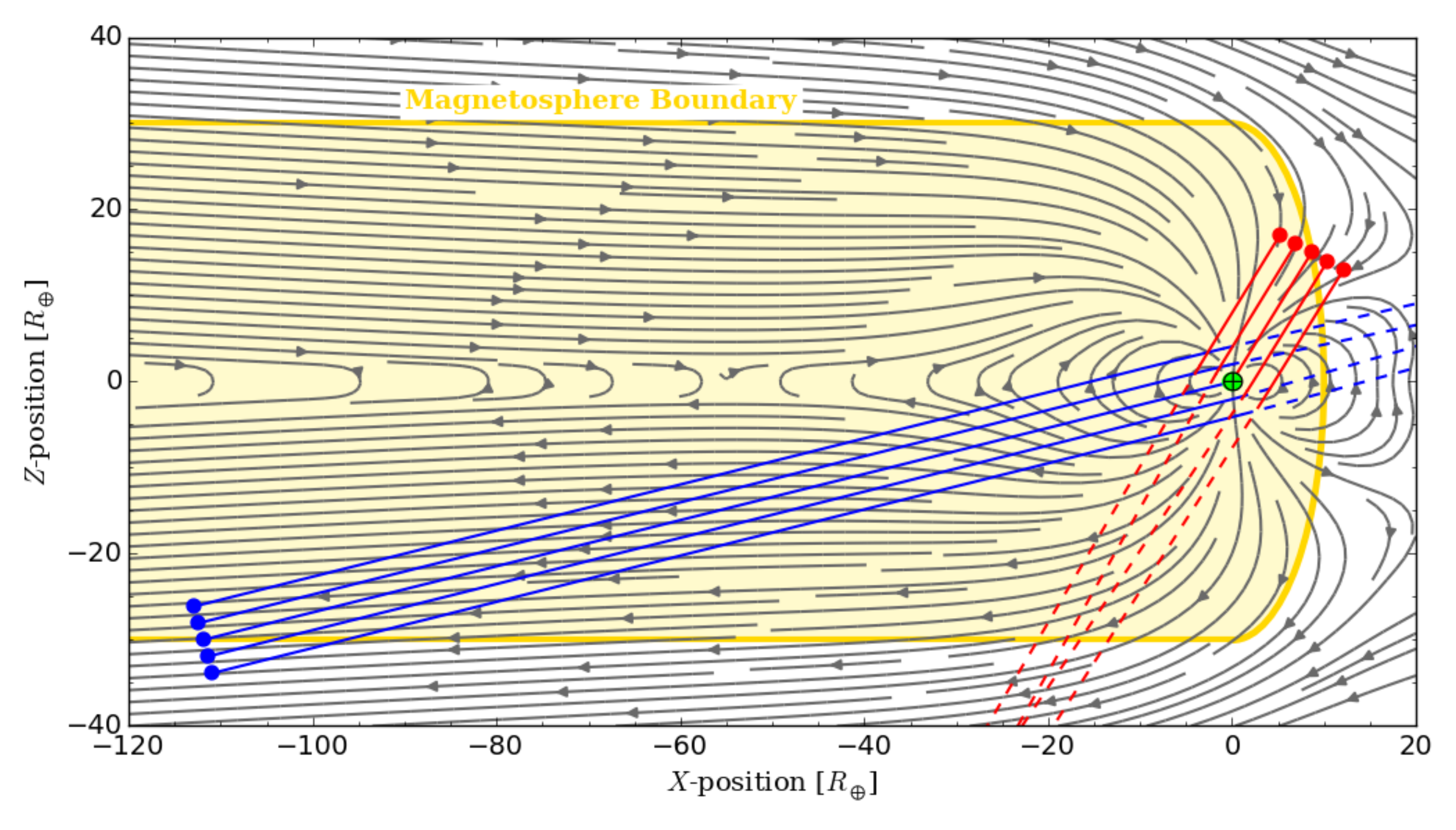}
		\label{fig:MagnetoMultiple}}
	\caption{Sample dust grain trajectories within the Magnetosphere.  The magnetic field lines are shown with grey arrows, the Earth is shown (to scale) by a green $\oplus$ at the origin, and the boundary of the Magnetosphere is shown with a yellow line (NOTE:  Magnetic field lines outside of the boundary were not used and can be ignored).  Dust grain trajectories are shown in red or blue with the initial locations indicated by dots, the incoming trajectory indicated by solid lines, and, after closest approach to Earth, the outgoing trajectory is indicated by dashed lines.  The initial dust grain parameters are:  $a = 0.2 \ \mu$m, ${\cal V} = 5$ V, and $v = 40$ km s$^{-1}$.  The left panel shows that individual dust grain trajectories initially aimed at Earth experience little deflection and impact Earth.  The right panel shows a dust swarm with grains initially travelling parallel to each other remain parallel until after passing Earth.
	\label{fig:MagnetoTrans}
	}
\end{figure*}

The Earth's Magnetosphere has a teardrop shape, with field lines on the day-side being compressed by 
solar wind pressure on the plasma frozen in the Magnetosphere, and being stretched on the night-side 
nearly parallel to one another.  The day-side edge is located $\sim$10 $\RE$ with a field strength about 
twice that of the dipole value \citep[\S 6.3.2,][]{kiv95}.  The night-side tail extends out to $\sim$1000 $\RE$ 
with a radius of $\sim$30 $\RE$ \citep[\S 9.3,][]{kiv95}.  It reaches asymptotically a field strength
$B_{X0} \approx 100 \ \mu$G \citep{slav85} and has a current sheet half-height of $H = \RE/2$ \citep{tsy89}.  
We use the Magnetosphere approximation from \citet{kats87}; this model is a superposition of a 
dipole field ($\vec{B}_{\rm dipole}$) near Earth \citep{drag65} and an asymptotic sheet ($\vec{B}_{\rm tail}$) 
for the magnetotail region \citep{wag79}.  This approximation does not include the inclination of the dipole field 
to the orbital plane but, given the motion and flipping of the magnetic poles, this approximation should suffice for examining 
general properties.  We assume a magnetic dipole strength based on the equatorial surface value of 
$M \approx 1 \ {\rm G} \ \RE^{3}$, and assume that the tail magnetic field normal component
$B_{Z0} \approx 0.06 B_{X0}$ \citep{slav85}.  When we solve the equations of motion (Equation (\ref{eq:eqnmotion}) adding the Lorentz force due to the Magnetosphere, $\vec{F}_{\rm mag, \oplus}$, and Earth's gravity, $\vec{F}_{\rm grav, \oplus}$) for a 
charged particle in a magnetic field starting at various locations at the edge of the Magnetosphere moving towards the Earth, 
we find deflections are $\lesssim 3$ arcmin when using our fiducial grain values.  Like the IMF, the grains show uniform deflections 
passing through the Magnetosphere, suggesting that direction information of the grains' source would be retained to within $10$ arcmin, 
and that spatial and temporal dilutions/concentrations of the \fe60 signal can be ignored, see Figure \ref{fig:MagnetoTrans}.

\subsection{Upper Atmosphere Distribution}
\label{subsect:atmodistro}

\begin{figure*}[t]
	\centering
	\subfigure[]
		{\includegraphics[width=\textwidth]{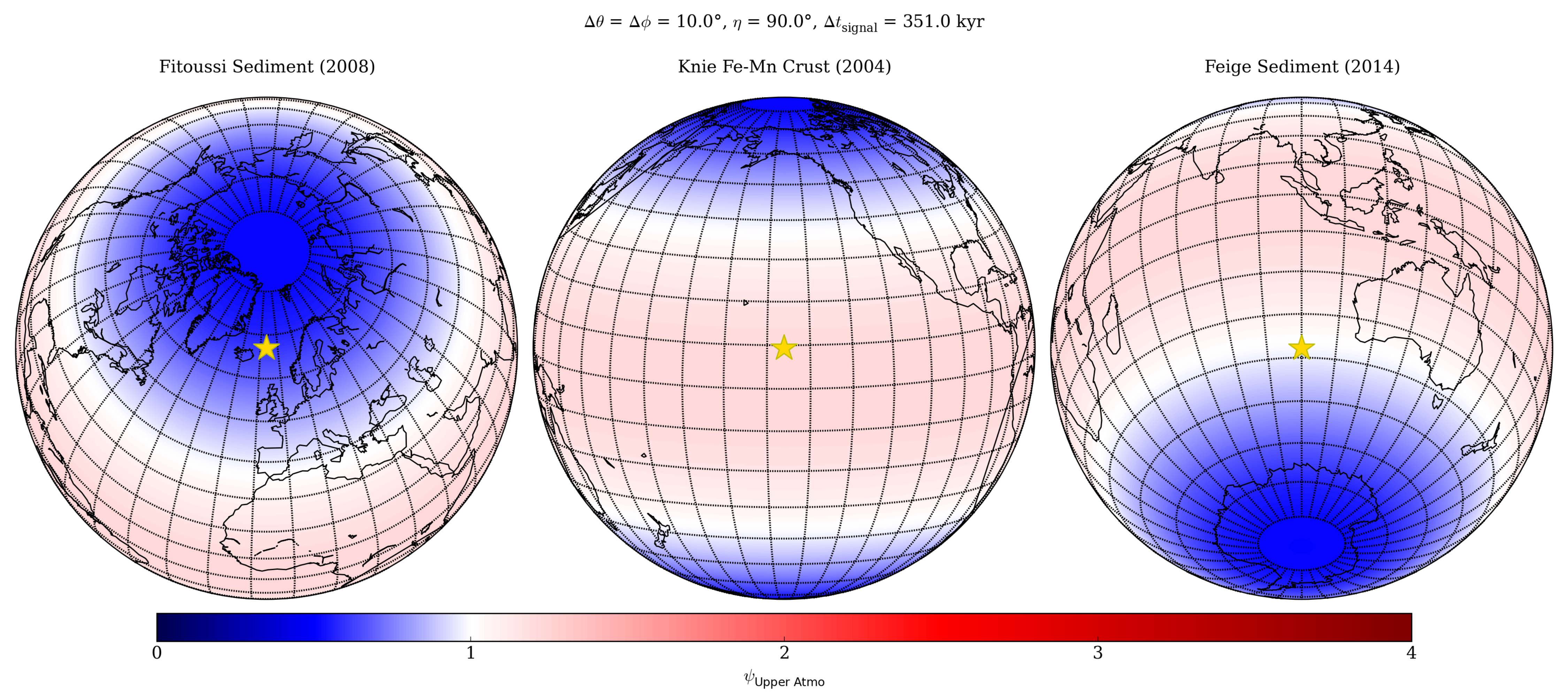}
		\label{fig:eta90}} \\
	\subfigure[]
		{\includegraphics[width=\textwidth]{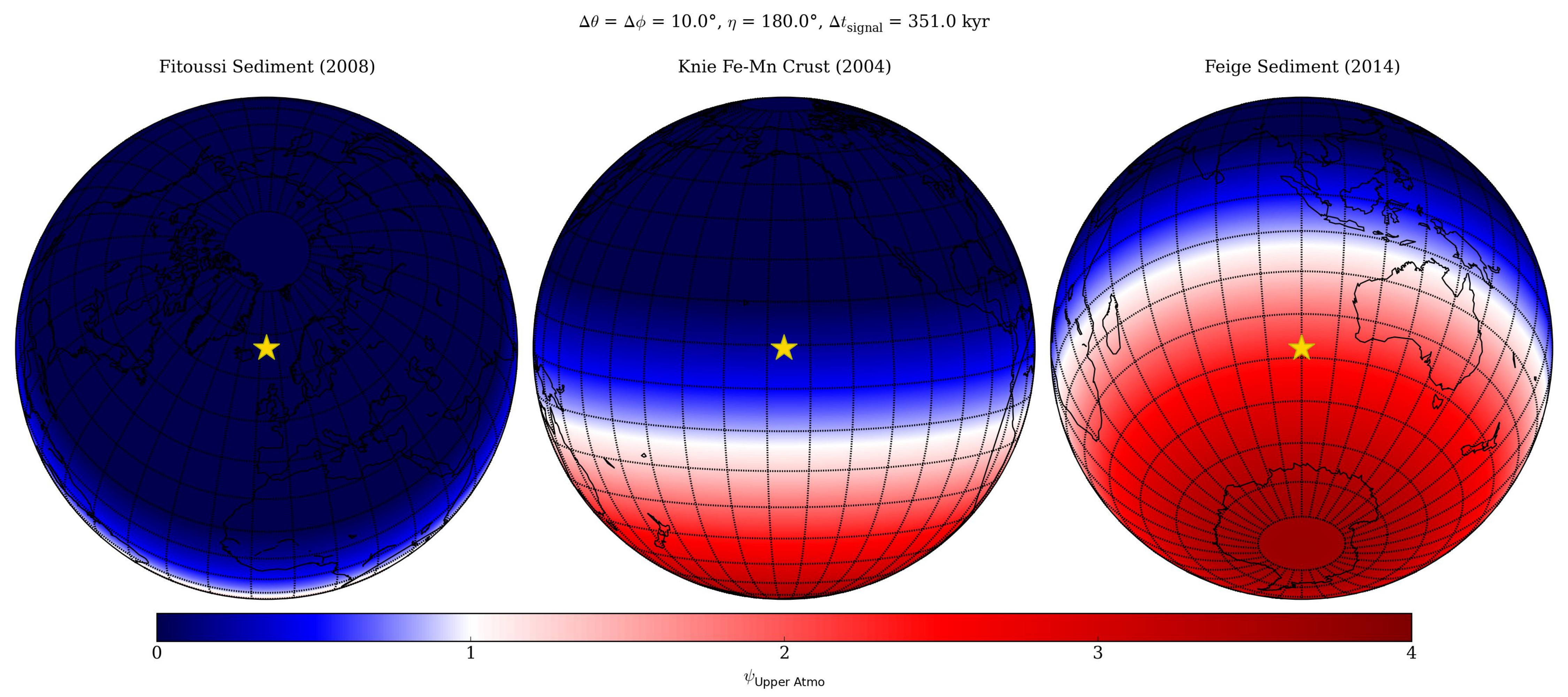}
		\label{fig:eta180}}
	\caption{Sample values of the distribution factor, $\psi$, as a function of the arrival angle, $\eta$ at the top of the atmosphere.  
	As $\eta$ increases from $\eta = 0^{\circ}$, the distribution changes from a northern concentration 
	to an equatorial concentration at $\eta = 90^{\circ}$.  The sampling locations are shown as yellow stars in the centers of the figures.  
	Note that, regardless of the value of $\eta$, the equator always receives some flux.  It should be noted that the plotting program used to make these figures automatically smooths the transition from grid to grid, making the figures appear of higher resolution than actually calculated.  However, based on the latitudinally-averaged values, the grid-to-grid transitions are, in fact, smooth, and the appearance shown in the figure is accurate.
	}
	\label{fig:distro}
\end{figure*}

\begin{figure*}[t]
	\centering
		\includegraphics[width=\textwidth]{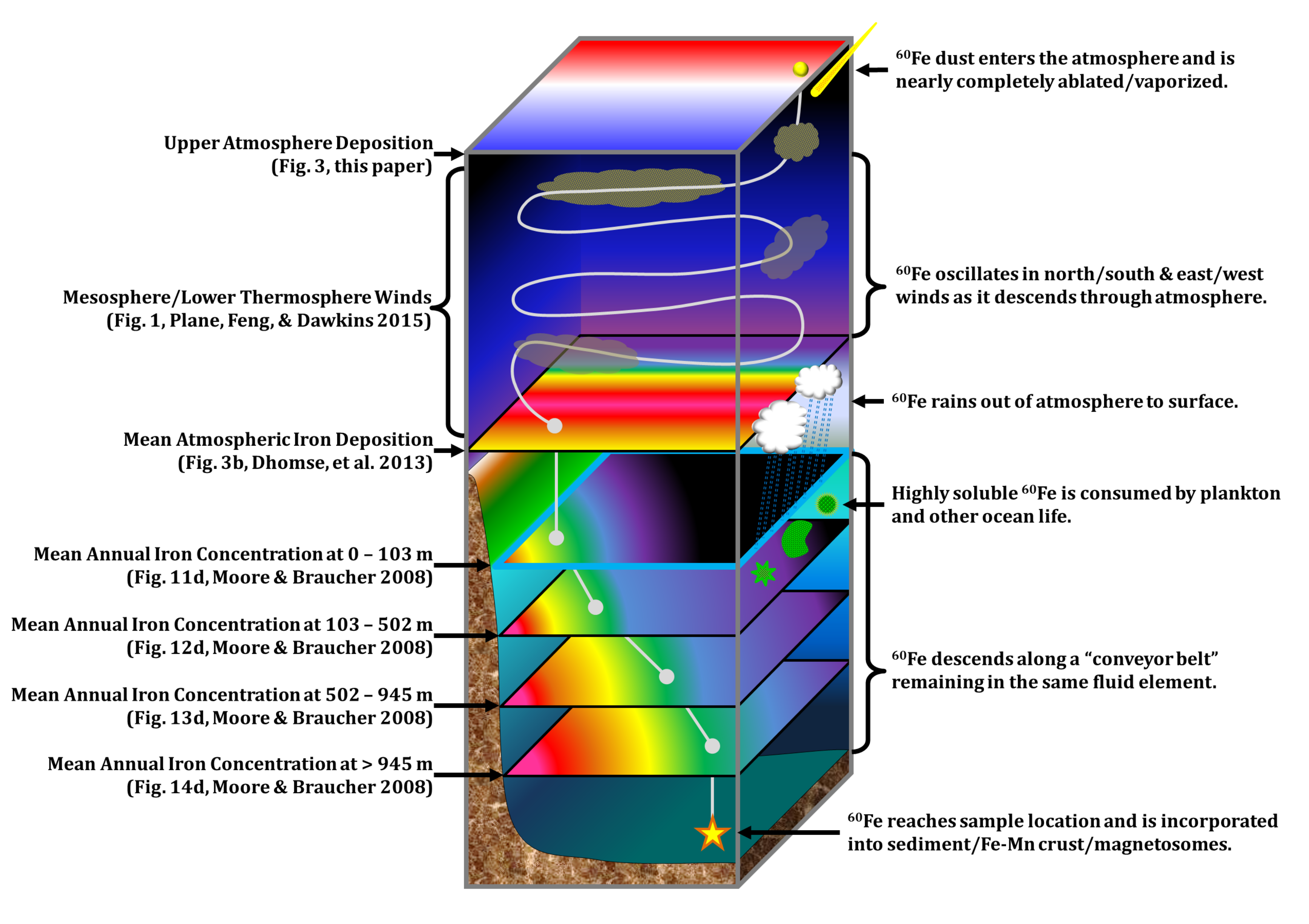}
		\caption{Schematic of \fe60 passage through the atmosphere to the ocean floor.  This diagram summarizes the processes and assumptions
		outlined in \S \ref{subsect:atmodistro}-\ref{subsect:water}.  On the left side of the diagram, the relevant references 
		used in tracking the \fe60 material's passage are given, and on the right side, the main processes acting on the \fe60 material are described.  Color gradients indicate concentration gradients of iron and mimic those found in the source figures referenced at left.  For this schematic, the colors of the gradients do not have specific values associated with them, but show how the referenced figures relate to one another.  A SN dust grain containing \fe60 enters at the top of the schematic, is vaporized, and the \fe60 vapor descends through the atmosphere until it is rained out to the surface.  This surface location is given in Fig. 3b, \citet{dhom13}, and the \fe60 material will enter the ocean fluid element at the corresponding location in Fig. 11d, \citet{moor08}; in this schematic, the fluid element has a ``green'' concentration.  The \fe60 remains in this ``green'' fluid element as it descends; the location of the ``green'' fluid element in Fig. 14d, \citet{moor08} will correspond to the sampling location on the ocean floor.  To determine the amount of wind and water deflection, we follow the \fe60 material's path backwards from the sampling location, along the fluid element with the associated iron concentration to the surface, and to the accompanying location on Fig. 3b, \citet{dhom13}.
		}
	\label{fig:windwater}
\end{figure*}

Once the SN dust has passed through the IMF and Magnetosphere, it impacts the upper atmosphere 
(generally at $\sim$100 km in altitude, see \S \ref{subsect:wind}).  Because the IMF and Magnetosphere show little deflection, we expect 
a relatively coherent, nearly plane-wave flow of incident dust onto Earth's upper atmosphere.  Once the grains reach Earth, they will 
impinge onto Earth's cross-section facing the dust wave.  The upper atmosphere distribution will depend on Earth's rotation 
and precession and the angle of arrival of the dust (see Figure \ref{fig:distro}).  To find the dust distribution in the 
upper atmosphere where the SN material impacts and before it begins to pass through the rest of the atmosphere, 
we approximate the Earth as a perfect sphere that rotates about the $z$-axis.  We divide the surface of the Earth
into sectors of angular size $\Delta \theta \times \Delta \phi$, with $\theta$ and $\phi$ analogous to latitude and 
longitude, respectively.  Because the duration of the SN dust storm is likely to be long ($\Delta t_{\rm signal} \sim 100$ kyr), 
we include Earth's axial precession ($\Delta t_{\rm precessional} = 26$ kyr).  We ignore nutation of Earth's axis,
since it is small ($\sim$arcseconds) compared to the Earth's inclination ($\alpha \approx 23.3^{\circ}$).  
Because the SN progenitor is far away ($D > 10$ pc), we assume the direction of the particle flux 
does not change with time and its intensity is uniform, so we ignore Earth's change in position through its orbit.  
We also assume that  the SN dust intensity varies with time according to the saw-tooth pattern used in \citet{fry15}:  
the initial flux ($\mathbb{F}_0$) starts at a maximum and decreases linearly to 0 at $t = \Delta t_{\rm signal}$.

In order to determine the fluence received at a given location on Earth, we use a series of 
coordinate transformations from the Earth's surface/terrestrial (unprimed) frame to the propagating shock wave/interstellar ($''''$) frame.  
For a detailed description of our transformations, see Appendix \ref{app:areacalc}.

Our simulations were run assuming a SN signal duration of $\Delta t_{\rm signal} = 351$ kyr \citep[the approximate 
expected duration for an ECSN,][]{fry15}.  
Because $\Delta t_{\rm signal} > \Delta t_{\rm precession}$, the model showed little dependence on the signal duration 
after the first precession cycle (terrestrial models were run for the entire SN signal width:  351 kyr; lunar models were run 
for four precession cycles:  74 yr).  The same is true for a constant flux profile versus a saw-tooth profile.  
Because the model includes two vastly different 
time scales (precessional and daily), we used two different time steps.  The precessional time steps were made
when the precession progressed by an angle $\Delta \phi/2$.  In other words:
\beq
\Delta t_{\rm precessional~step} = \left( \frac{26 \ {\rm kyr}}{360^{\circ}} \right) \left( \frac{\Delta \phi}{2} \right) \, .
\label{eq:timeprec}
\eeq
At each precessional time step, the model is run for one daily rotation, with the daily time steps made
when the daily rotation progresses by an angle $\Delta \phi/2$, or:
\beq
\Delta t_{\rm daily~step} = \left( \frac{86400 \ {\rm s}}{360^{\circ}} \right) \left( \frac{\Delta \phi}{2} \right) \, .
\label{eq:timeday}
\eeq
Precession still occurs during the daily time steps, but the effects of the daily rotation dominate.  
As we ran our model, the various angles $\eta$ represent different arrival directions from the source of the 
\fe60 signal as measured from the Ecliptic North Pole.  Because of Earth's precession and rotation, 
these possible directions form a ring of constant Ecliptic latitude.

Figure \ref{fig:distro} shows sample results for our upper atmosphere distribution model, and we can see for the $\eta = 90^{\circ}$ case, there is a nearly isotropic distribution of particles onto the entire atmosphere; $\psi_{\rm Upper~Atmo,~\eta=90^{\circ}} \in [0.5, \ 1.2]$.  As $\eta$ increases to 180$^{\circ}$, the North Pole becomes increasingly depleted ($\psi \rightarrow 0$), and the South Pole becomes increasingly saturated (the $\eta \in [0, \ 90^{\circ})$ case mirrors this result).  At $\eta = 180^{\circ}$, the saturation reaches a maximum; $\psi_{\rm Upper~Atmo,~\eta = 180^{\circ}} \in [0, \ 3.7]$.

We see in Figure \ref{fig:distro} that the arrival distribution of SN material is uniform
across longitudes (i.e., constant at a fixed latitude).  This arises primarily due to the daily rotation, with 
some additional smearing due to precessional rotation.  Conversely, the distribution of SN dust on the upper atmosphere 
is strongly nonuniform across latitudes.  The latitude gradient largely reflects the direction of the SN itself, 
with some smearing due to precession.  As we will see, the fate of this SN signature is very different for the Earth and Moon.

\subsection{Wind Deflection}
\label{subsect:wind}

Interstellar dust containing \fe60 could be subject to two types of wind effects:  
initial deflection through the atmosphere and subsequent transplantation from a landmass into the ocean.  
Since the solar wind has little influence on the SN dust grains, they would enter Earth's atmosphere 
at approximately the same speed they entered the Solar System: $v_{\rm SN \ grains} \approx 40-100$ km s$^{-1}$.  
Although this is faster than typical meteoritic dust infall velocities, we would expect SN dust to be ablated 
at similar altitudes to meteoritic dust because both are traveling supersonically relative to the 
surrounding air and the stopping distance is independent of the initial velocity: in the supersonic limit, 
the $e$-folding stopping distance for dust grains is independent of their initial velocity \citep{murr04}.  
This implies that the SN dust grains would come to rest relative to the atmosphere in the upper mesosphere/lower thermosphere 
\citep[MLT, $\sim$$90-115$ km above sea level,][]{feng13}.  However, because of their high velocities, 
we would expect the SN grains to be completely ablated upon impact with the atmosphere,
and thus vaporized.  At this point, the SN \fe60 vapor would descend through the atmosphere (see Figure \ref{fig:windwater}).

We expect that the SN dust grains and meteoritic dust grains would be similar in size ($a \sim 0.1-1 \ \mu$m),
so their ablation and fragmentation properties would also be similar.  The SN grains would be ablated at altitudes
similar to where meteoric grains are ablated, and both would descend through the atmosphere in a similar manner.
Their compositions (iron oxides and silicates) are identical, so both SN and meteoritic materials would experience 
similar chemical reactions in the atmosphere.  Because of these similarities, we use the extensive work already 
accomplished on meteoric smoke particles \citep[e.g.,][and references therein]{plane15}.

Once delivered to the MLT, the SN material would sediment out to the surface over the course of 
$4-6$ years \citep{dhom13}.  As noted in \S \ref{subsect:atmodistro}, because of Earth's rotation and precession, 
the upper atmosphere distribution forms bands of uniform fluences across lines of latitude.  
Since zonal (east-west) deflection would not affect that pattern, we focus on deflections due to meridional (north-south) winds.  
In the MLT, meridional winds are of the order $v_{\rm MLT \ winds} \sim 10$ m s$^{-1}$ and can be several 
orders of magnitude greater than the vertical component \citep[Figure 1,][]{plane15}.  
These winds could drive the SN material from one pole to the other within a few days while
descending only a few kilometers.  An example of this movement was the plume from the launch of 
STS-107 on January 16, 2003: within $\sim$80 hr the plume had traveled from the eastern coast of Florida to the 
Antarctic \citep{nici11}.  Downward transport through the mesosphere-stratosphere-troposphere occurs mainly
in the polar regions: this leads to a semi-annual oscillation of
meteoritic smoke particles from pole to pole that would effectively isotropize 
(or at least randomize) the distribution of incoming SN material in the mesosphere.

{\renewcommand{\arraystretch}{1.5}
\begin{table*}[t]
\begin{center}
\caption{Summary of Dust Grain Transit Through Solar System to Ocean Floor}
\label{tab:summary}
	\begin{tabularx}{\textwidth}{ l C c c } \hline \hline
			Region					& Primary Influences 						& Residence Time		& Characteristic Region Boundary* 	\\
			Interplanetary			& IMF, Solar Radiation/Gravity				& $\sim 10-12$ years	& $100-150$ AU				\\
			Magnetosphere			& Terrestrial Magnetic Field				& $\lesssim2$ days		& $10-1000 \ \RE$			\\
			Upper Atmosphere (MLT)	& Collisional Drag/Ablation \& Strong Winds	& $4-6$ years			& $90-115$ km				\\
			Troposphere				& Rain, Wind								& $1-2$ weeks			& $\sim 10$ km				\\
			Surface:				& 											& 						& 							\\
			~~~~~- Land				& Transplantation by Surface Winds			& indefinite			& $\sim$ km			\\
			~~~~~- Water			& Biological Uptake, Ocean Currents			& $100-200$ years		& $\sim 4$ km				\\
			Ocean Floor				& Biological Transplantation/Disturbance	& indefinite			& $\sim$ m					\\ \hline
			\multicolumn{4}{ p{\textwidth} }{* - Boundary distances are measured from Earth's surface except for the Water and Ocean Floor regions which
			are measured from the ocean floor.}
	\end{tabularx}
\end{center}
\end{table*}}

In addition, the vaporized SN \fe60 would be highly soluble and would combine with sulphates as it descended
through the stratosphere \citep{dhom13}.  This means the SN material would be readily incorporated into 
clouds when it finally reaches the troposphere \citep{saun12}.  Because the SN \fe60 would behave
similarly to meteoritic iron, we can use simulations of the meteoritic smoke particles
to find the final distribution of SN \fe60 
at the surface.  \citet{dhom13} studied the transport of \pu238 through the atmosphere and later applied their model to 
iron deposition, finding the distribution over the entire Earth, with asymmetries in the mid-latitudes due to the 
stratosphere-troposphere exchange \citep[see Figure 3b,][]{dhom13}.

After descending through the atmosphere, it is possible for interstellar dust grains that have fallen through the atmosphere 
and been deposited on land to later be picked up by wind again, carried to the ocean, and be deposited there.  
This process of dust transplantation (also called aeolian dust), could lead to an enhancement of 
\fe60 levels in ocean samples.  In the case of our studied samples, however, this should not be an issue.  
Based on a study by \citet{jick05}, aeolian iron dust deposits are low in the area of the \citet{knie04} and 
\citet{feige14} samples.  While slightly higher than the other locations, the \citet{fit08}
sample should not be affected because of where the dust was transplanted from.  In the case of the Fitoussi sediment samples, 
the material will be transplanted from equatorial regions (e.g., to the Sahara, Arabian, and Gobi deserts), but as described 
in \citep[see Figure 3b,][]{dhom13},  these areas will receive very little SN material so we would expect the transplanted 
dust to contain a negligible amount of SN \fe60.\footnote{Moreover, our use of an upper limit 
for the Fitoussi sample should allow for any transplantation enhancement.}  Therefore, for the purposes of this paper we ignore dust transplantation, but future studies should consult \citet{jick05} to check if transplantation is an issue.

{\renewcommand{\arraystretch}{1.5}
\begin{table*}[t]
\begin{center}
\caption{Predicted Fluence Ratios for Uptake Values}
\label{tab:results}
	\begin{tabularx}{\textwidth}{ C C C C C } \hline \hline
			\multicolumn{5}{ c }{$\psi_{\rm Knie} = 0.43$, $\psi_{\rm Fitoussi} = 0.14$, $\psi_{\rm Feige} = 1.4$} \\ \hline
			Fluence Ratios										& Observed 			& High Uptake 	& Medium Uptake 	& Low Uptake	\\
			${\cal F}_{\rm Fitoussi}/{\cal F}_{\rm Knie} =$ 	& $0. \ (<70.9)$	& 0.33 			& 0.65 				& 3.3  			\\
			${\cal F}_{\rm Fitoussi}/{\cal F}_{\rm Feige} =$	& $0. \ (<7.04)$	& 0.11 			& 0.11 				& 0.11			\\
			${\cal F}_{\rm Feige}/{\cal F}_{\rm Knie} =$ 		& $10. \pm 6$ 		& 3.3 			& 6.5 				& 33 			\\ \hline
	\end{tabularx}
\end{center}
\end{table*}}

\subsection{Water Deflection}
\label{subsect:water}

As mentioned in \S \ref{subsect:wind}, the SN material would be highly soluble due to its complete ablation in the MLT.  
This means that when it reached the ocean it would be incorporated readily into organisms, particularly 
phytoplankton \citep{boyd10}.  In many locations, the availability of iron is the limiting factor for
phytoplankton growth \citep[Figure 7,][]{moor04}.  In locations where there is an abundance of iron (i.e., 
high concentrations of soluble iron, most likely due to meteoric or aeolian sources, and iron is not the limiting element), 
the residence time for iron is very short ($\sim$days and months), but in locations of lower abundance of Fe, the residence time is longer 
($\sim$$100-200$ years) \citep{brul94, croo04}.  In either case, these residence times are much less than the 
ocean circulation time ($\sim$1000 years).

When quantifying the distribution of iron as it descends in the ocean, a number of considerations need to be included,
not only the initial location of iron, the water velocity and its depth, but also the complexation of iron with 
organic ligands, the availability of other nutrients such as phosphates and nitrates, seasonal patterns, 
ocean floor topography, and the amount of light exposure.  Several studies have examined iron cycling in the 
ocean \citep[see e.g.,][]{lefe99,arch00,pare04,dutk05,dutk12}.  However, all of these studies examine the 
total iron input into oceans, the dominant source being aeolian dust which is highly insoluble, rather than 
meteoritic sources that are highly soluble but account for only $10^{-4}$ of the total iron input mass \citep{jick05,plane12}.

A more recent study by \citet{moor08} examined the global cycling of iron and updated the 
Biogeochemical Elemental Cycling (BEC) ocean model, resulting in an improved model 
that showed better agreement with observations.  As part of this study, \citeauthor{moor08} simulated the 
concentrations of dissolved iron at varying ocean depths; of particular interest are the simulations of ``Only Dust'' 
inputs \citep[see Figures 11d, 12d, 13d, and 14d, ][]{moor08}.  While the dust used in the simulation is primarily 
from an aeolian source, it acts similarly to meteoric dust (or interstellar SN dust) upon reaching the ocean.  
Since the residence time of iron is much less than the ocean circulation time, we can approximate the 
ocean currents as ``conveyor belts'', moving different concentrations of iron to different areas of the ocean, 
but not significantly altering the concentration of a fluid element as it descends.  With this assumption, 
we can find a first-order, initial location of the dust input by looking at the iron concentration over each \fe60 sampling
location in the lowest depths \citep[Figure 14d, ][]{moor08} and following it back to its source on the surface 
\citep[Figure 11d, ][]{moor08}.  With this initial location, we can use the meteoric dust distribution from 
Figure 3b, \citet{dhom13} at that location to find the relative fractions of SN \fe60 that would eventually reach the
sampling locations (see Figure \ref{fig:windwater}).  Using this method, we would expect the material deposited in the Knie crust sample to have 
originated from the Sea of Okhotsk off the northern coast of Japan, the Fitoussi sediment sample to have 
originated from the northwestern coast of Africa near the Strait of Gibraltar, and the Feige sediment sample
to have originated between the southern tip of Africa and Antarctica.

\section{Results}
\label{sect:results}

We present results of the surface distribution patterns for both the Earth and Moon.

\subsection{Terrestrial \fe60 Distribution}
\label{subsect:terrdistro}

Comparing the various influences on the SN material, we find that the influence of the atmosphere 
(in particular, the MLT) would have been the greatest determining influence on the distribution of SN 
material at the sampling sites.  The IMF, Magnetosphere, and water currents can deflect SN material, 
but these effects are small in scale and/or systematic in nature. Moreover, while the arrival angle, $\eta$, 
certainly causes global variations in received fluence, these variations would have been completely lost 
as the SN material descended through the MLT.  A summary of a SN dust grain's transit is given Table \ref{tab:summary}.

Therefore, because motions in the MLT remove information of the original SN dust's direction, the terrestrial \fe60 distribution provides no useful clues as to the SN origin on the sky.  This is not to say, however, that the terrestrial \fe60 distribution should be
uniform.  Rather, the surface pattern reflects terrestrial transport properties.

To find the distribution factors, $\psi$, we use the annual mean iron deposition rates from \citet{dhom13} 
corresponding to the initial locations identified using \citet{moor08} and the model's total global input of 
$27 \ {\rm t \ day^{-1}} \Rightarrow \mathbb{F}_{\rm global} = 0.35 \ {\rm \mu mol \ m^{-2} \ yr^{-1}}$.  
This yields distribution factors at the sampling locations of:  $\psi_{\rm Knie} = 0.15/0.35 = 0.43$, 
$\psi_{\rm Fitoussi} = 0.05/0.35 = 0.14$ and $\psi_{\rm Feige} = 0.5/0.35 = 1.4$.  These results are notable,
first because they are not equal to unity, and secondly because they are still within an order of magnitude of unity.  
This means that if we compare the isotropic and anisotropic distributions in Equation (\ref{eq:fluence}), we find
that $D_{\rm anisotropic}/D_{\rm isotropic} \approx \sqrt{\psi}$. Therefore, based on our estimated distribution factors, 
a SN distance calculated assuming an isotropic distribution would still be within of an order of magnitude of a 
full calculation including distribution effects.  Using these distribution values and the uptake values for each case, 
we can compare the fluence ratio predictions with the observed values, as shown in Table \ref{tab:results}.

\begin{figure*}[t]
	\centering
	\subfigure[]
		{\includegraphics[width=0.45\textwidth]{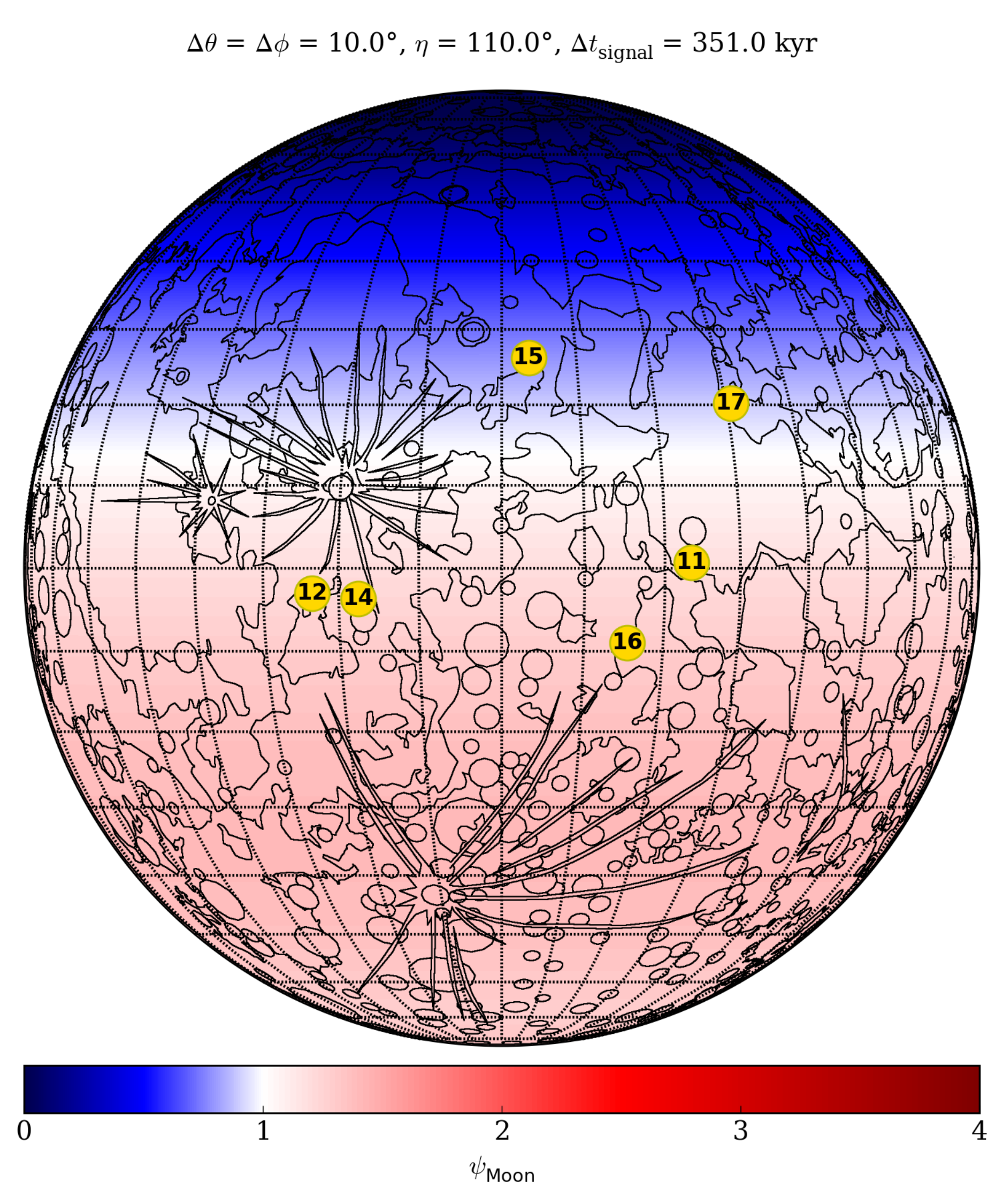}
		\label{fig:MoonScoCen}} ~	
	\subfigure[]
		{\includegraphics[width=0.45\textwidth]{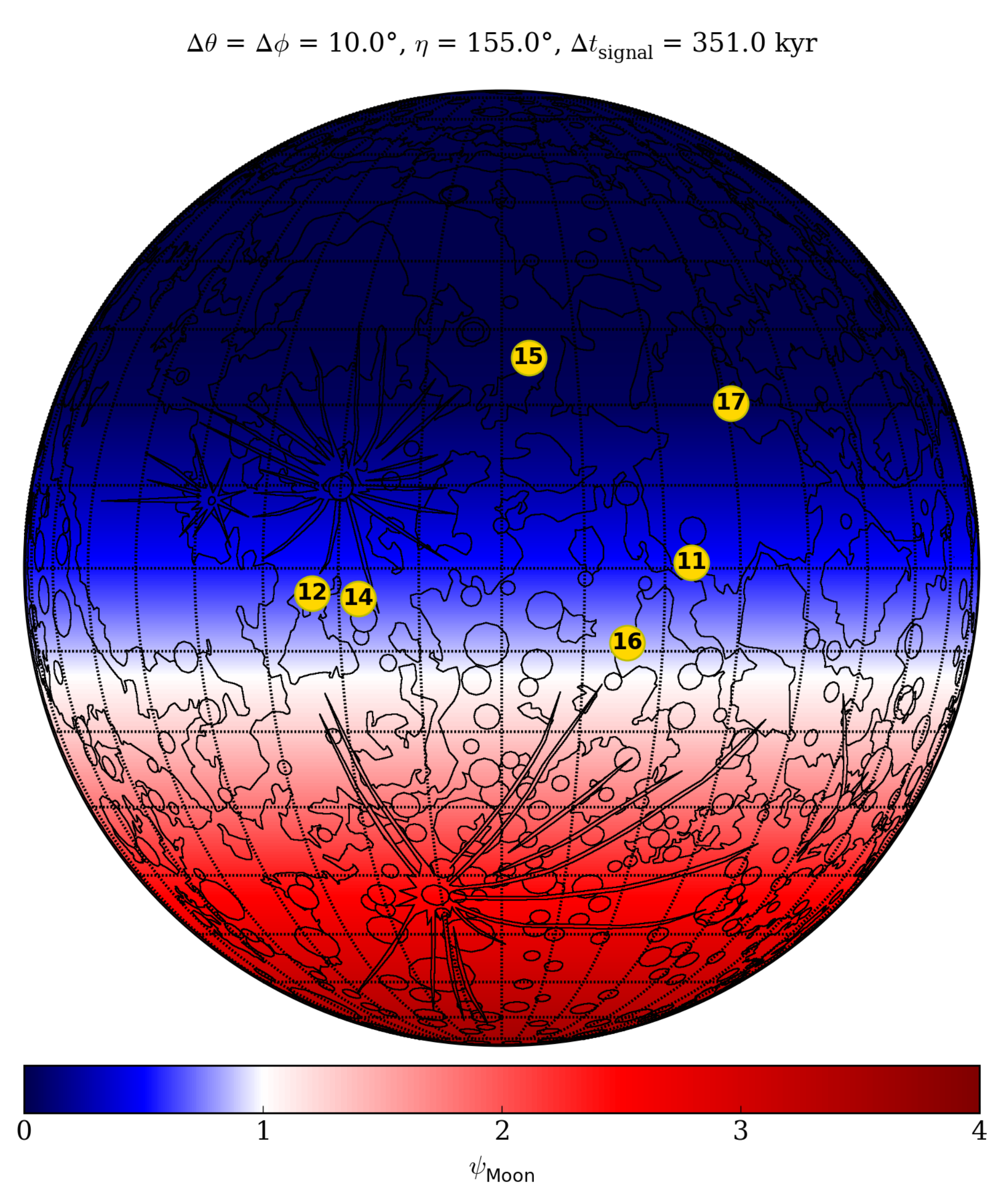}
		\label{fig:MoonTucHor}} \\
	\subfigure[]
		{\includegraphics[width=\textwidth]{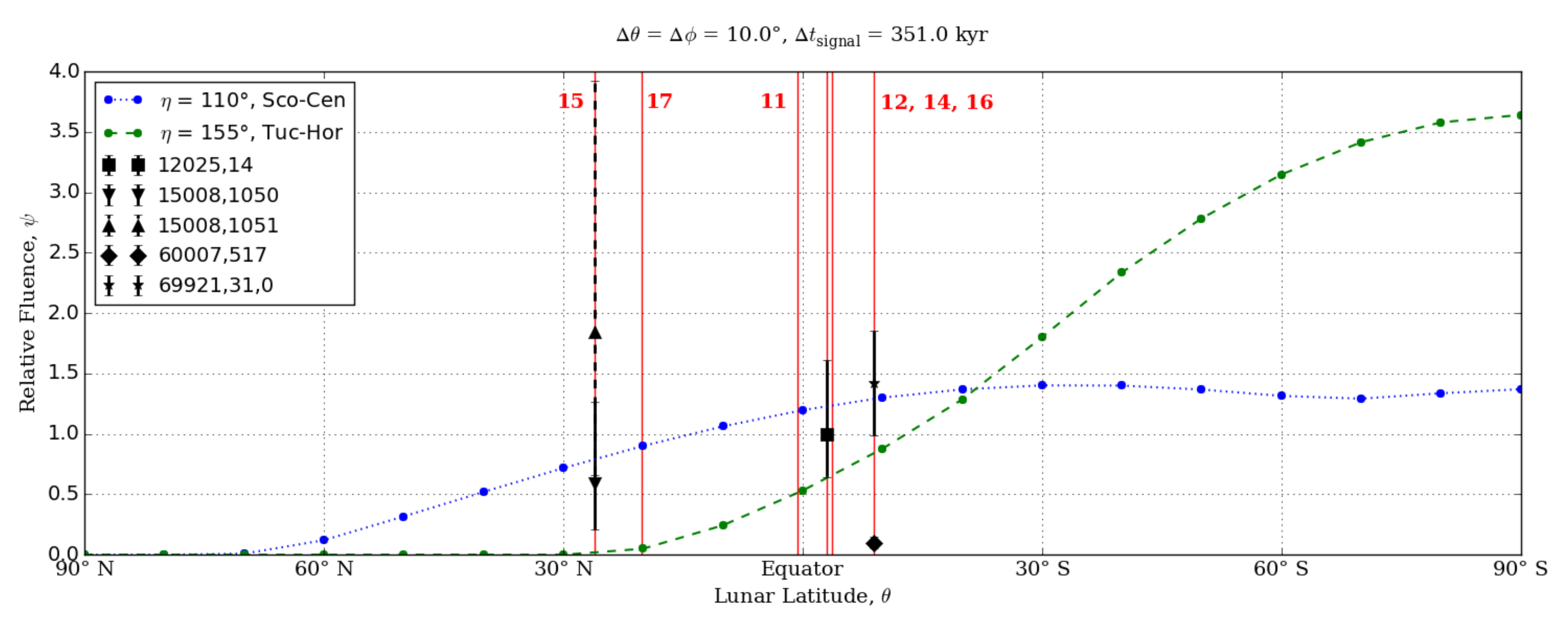}
		\label{fig:LunarAve}}
	\caption{Sample predicted values for the lunar distribution factor, $\psi_{\rm Moon}$.  For SN material arriving from $\eta = 110^{\circ}$, corresponding to a SN in the Sco-Cen region (top, left panel), and $\eta = 155^{\circ}$, corresponding to a SN in the Tuc-Hor region (top, right panel).  Apollo landing sites are highlighted by the numbered, yellow circles in the upper panels and vertical, red lines in the lower panel \citep{dav87, dav00}.  The \citet{fimi16} measurements are shown in black; of note are the large error ranges particularly for the 15008 samples.  Like Figure \ref{fig:distro}, the plotting program smooths grid-to-grid, however, the lower panel shows the latitudinally-averaged relative fluences for the top panels.  The actual model averages are shown with data points, and the connecting lines show a smooth, nearly sine-function profile.  The lunar background diagram used with permission from Steven Dutch, University of Wisconsin at Green Bay.
	\label{fig:moondep}
	}
\end{figure*}

\subsection{Lunar \fe60 Distribution}
\label{subsect:lunardistro}

In contrast to the Earth, the airless and dessicated Moon will introduce none of the atmospheric and oceanic transport
effects that influence the terrestrial \fe60 deposition on the ocean floor.  In particular, lunar deposition of SN debris will not
suffer the large smearing over latitudes that plague material passing through the Earth's MLT.  Consequently, the lunar
distribution of SN debris holds to hope of retaining information about the SN direction.

Like Earth's upper atmosphere, dust grains impacting the lunar surface would be deflected 
$\lesssim 1^{\circ}$ from their passage through the Solar System, but the lunar deposits would not be further shifted 
by wind/water.  Because atmospheric and ocean effects can be ignored, the SN directionality will be preserved.  
We can adapt the method for finding upper atmosphere deposition (\S \ref{subsect:atmodistro} and 
Appendix \ref{app:areacalc}) by using lunar parameters (daily period:  27 days, precessional period:  19 years, 
inclination angle:  6.7$^{\circ}$).  The deposition forms a banded pattern like that shown in Figure \ref{fig:moondep}.
We again see that the distribution is uniform across longitudes due to lunar
rotation, but a latitude gradient persists and reflects the SN's direction, smeared somewhat by precession.
The upper left panel (Figure \ref{fig:MoonScoCen}) assumes $\eta = 110^{\circ}$, corresponding to a SN in the Sco-Cen region, 
and the upper right panel (Figure \ref{fig:MoonTucHor}) is based on a source with $\eta = 155^{\circ}$, 
corresponding to a SN in the Tuc-Hor region.

\citet{fimi16} recently published measurements of lunar \fe60 collected during the Apollo Moon landings.  The measurements cover a range of depths, but an impactor will only penetrate to a depth on order with its diameter ($\sim \mu$m for our SN grains), so we compare only the fluences of the shallowest samples \citep[this will allow for some minor gardening as noted by][and references therein]{fimi16}.  We calculated fluences as outlined in \S \ref{subsect:FiSamp}, adjusted for 10\% cosmogenic (i.e., cosmic ray-produced) \fe60, and plotted the results against the expected Sco-Cen and Tuc-Hor relative fluences in Figure \ref{fig:LunarAve}.  Since we do not know the actual fluence for an isotropic distribution on the Moon, we scaled the fluences to the 12025,14 sample fluence.  We see that it is not yet possible to differentiate between a Sco-Cen or Tuc-Hor source because the samples were drawn near the lunar equator and the large uncertainties in the measurements (we also note that a future, more detailed examination should address the effects of regolith composition, gardening, and impactor penetration depth).  The uncertainties are the result of low-number statistics \citep[the two plotted 15008 values are from a total of four events, see][Supplemental Information]{fimi16}, but continued study will further refine these values.

\section{Conclusions}
\label{sect:conclusions}

After examining the major influences on SN material as it passes through the Solar System to the bottom of the ocean, 
we find that previous works' assumption of an isotropic terrestrial distribution of SN material was rather na{\"i}ve but,
based on our results, this assumption nevertheless yields calculated distances within an order of magnitude of a 
full calculation incorporating a distribution factor.  The dominant influence on the final distribution of 
SN material deposited on the Earth is the atmosphere, specifically the MLT region, due to strong zonal and meridional winds.  
This means that the suggestion by \citet{fry15} that the direction of arrival is the dominant cause of differing fluence 
measurements is incorrect.  Whilst the angle of arrival of SN material can have drastic effects on the SN material's 
initial distribution in the upper atmosphere, these variations are completely masked as the material descends to the 
surface.

However, although the method outlined in \S \ref{subsect:atmodistro} may not be applicable to finding the final distribution 
on Earth, \fe60 measurements using lunar regolith could apply the method.  We indeed find that the lunar distribution of \fe60 
retains information about the SN direction.  Namely, lunar rotation and precession average over
longitudes but preserve a latitude gradient that peaks near the SN latitude.  The recent exciting detections of \fe60 
from Apollo soil core samples show a proof of principle that the Moon can act as a telescope pointing to the SN.  
As yet the data, clustered at the lunar equator, are too uncertain to cleanly discriminate the two putative star cluster origins
(Sco-Cen versus Tuc-Hor), but future measurements -- or ideally, a sample return mission from high and low lunar latitudes -- could 
identify one possibility.

Clearly there are a number of uncertainties and assumptions included in our examination.  
The fluence ratios have large error uncertainties ($\sim$50\%) or are simply upper limits.  
This is a by-product of the counting statistics in making the \fe60/Fe measurements, 
and future \fe60 measurements will better constrain these values.  The uncertainty in the value of 
$\ucrust$ further complicates the fluence ratios and, whilst most likely $\ucrust \in [0.1, \ 1]$, 
the use of sediment samples would be preferable since $\usediment \approx 1$ is much more certain.  
Lastly, the application of \citeauthor{moor08}'s updated BEC model to our SN \fe60 ocean transport has
some limitations.  Although it includes many of the relevant considerations outlined in \S \ref{subsect:water}, 
it focuses on aeolian dust sources of iron rather than meteoric sources, which have a different starting distribution.  
Additionally, the updated BEC simulations match observations better than the previous model, but still 
rely on observations primarily from the northern Pacific Ocean \citep{moor08} and underestimate the 
deep ocean iron concentrations.  Also, we used a general conveyor belt assumption of the movement of iron in 
\citeauthor{moor08}'s results, rather than following tracer particles to understand better any possible dilutions or 
concentrations.  Because generating an ocean model to track our SN \fe60 material as it descends in the oceans 
with all the relevant factors described above is beyond the scope of this paper, we attribute any deviations from 
our observed fluence ratios and our predictions to errors in modeling iron transport within the oceans.

Based on our results, we can duplicate the observed fluence ratios.  The predictions for the Medium case 
show good agreement with observations of all ratios. In the case of the ${\cal F}_{\rm Feige}/{\cal F}_{\rm Knie}$ ratio, 
the High and Low uptake values give ratios outside the error ranges and a factor $\sim$3 from the mean value.  
In addition, comparing the \citet{feige14} and \citet{wall16} calculation of $\ucrust \in [0.7, 0.17]$ assuming $\psi = 1$, 
we find good agreement with our Medium case and our calculated distribution factors.  Revisiting Equation (\ref{eq:kniefeige}), for 
\citet{feige14}/\citet{wall16}:
\beq
\frac{{\cal F}_{\rm Knie}}{{\cal F}_{\rm Feige}} = \frac{U_{\rm Knie} \psi_{\rm Knie}}{U_{\rm Feige} \psi_{\rm Feige}} = 
\frac{0.1 \cdot 1}{1 \cdot 1} = 0.1 \, ,
\eeq
and for this work:
\beq
\frac{{\cal F}_{\rm Knie}}{{\cal F}_{\rm Feige}} = \frac{0.5 \cdot 0.43}{1 \cdot 1.4} = 0.15 \in [0.7, 0.17] \, .
\eeq
Although it would be preferable to compare a sediment and crust sample drawn from the same place in the ocean to directly measure the crust uptake, 
this suggests that the \citet{feige14,wall16} $\ucrust$ values inherently include the distribution factors between sampling locations.

Moreover, using Equation (\ref{eq:fluence}), we see that changing the uptake also changes the calculated distance to the source for a given 
observed fluence; increasing the uptake $U$ increases the estimated distance, $D$, and conversely decreasing $U$ decreases $D$.  If we assume that the SN that produced the measured \fe60 occurred in a stellar group (as opposed to being the explosion of an isolated star), 
we can compare the distances implied by each of our cases and the locations of the two candidate groups
Sco-Cen and Tuc-Hor.  Adapting the conditions outlined in \citet{fry15} to include the distribution factor, $\psi$, 
we find for an ECSN, in the Medium Uptake case, the implied distance is:  $D = 46_{-6}^{+10}$ pc, which is 
consistent with the distance to Tuc-Hor ($\lesssim 60$ pc) but not with Sco-Cen ($\sim$130 pc).  Table \ref{tab:distances} summarizes the implied distances for our uptake cases.

{\renewcommand{\arraystretch}{1.5}
\begin{table}[t]
\centering
\caption{Implied Source Distances for Each Uptake Case}
\label{tab:distances}
	\begin{tabularx}{0.48\textwidth}{ C c c c } \hline \hline
	Case					& High 						& Medium					& Low						\\
	$\psi_{\rm Knie} = 0.43$& $\ucrust = 1.0$			& $\ucrust = 0.5$			& $\ucrust = 0.1$			\\
	$8-10$-$\msol$ ECSN		& $45_{-6}^{+10}$ pc		& $46_{-6}^{+10}$ pc		& $35_{-5}^{+8}$ pc			\\
	15-$\msol$ CCSN			& $61_{-8}^{+14}$ pc		& $64_{-8}^{+14}$ pc		& $47_{-6}^{+11}$ pc		\\
	9-$\msol$ SAGB			& $82_{-8}^{+13}$ pc		& $84_{-8}^{+13}$ pc		& $67_{-7}^{+11}$ pc		\\ \hline
	\end{tabularx}
\end{table}

Finally, with regards to the number of SNe producing the \fe60 signal, we find no process within the Solar System that could spread the deposition of a single SN signal to appear like that found by \citet{wall16}.  Such a process would need to allow concentrated \fe60 to pass fairly undisturbed, but delay diluted \fe60.  The only process to make such a distinction is ocean cycling, where the residence time for iron decreases when there is an overabundance of iron (\S \ref{subsect:water}), however, the delay is only $\sim 100$ years not the $\gtrsim 100~{\rm kyr}$ required to reproduce the \citet{wall16} measurements.  In addition, $\fe60/{\rm Fe} \lesssim 10^{-14}$ in the ocean, so any \fe60 of SN origin would have no appreciable effect on ocean iron abundance.  This suggests either there were multiple SNe as postulated by \citet{breit16} and \citet{wall16} or another process within the ISM or SN remnant is responsible for spreading the signal.

\acknowledgments
The authors would like to thank Nicole Riemer, Sandip Dhomse, and John M. C. Plane for the their enlightening 
discussions of aerosols in the atmosphere.
We are grateful to E. Mamajek for drawing our attention to and discussing his 
work on the Tuc-Hor group as a strong candidate for the SN site.
B.J.F. would like to give Ashley Orr special thanks for discussions that greatly improved
the title and presentation of this paper.  We would also like to thank our reviewer whose thoughtful and thorough
comments on the manuscript greatly improved this work.  The work of J.E. was supported in part by the European Research Council via the 
Advanced Investigator Grant 267352 and by the UK STFC via the research grant ST/L000326/1.

{}

\appendix

\section{A. Coordinate Transformations for Calculating Fluence onto a Sector}
\label{app:areacalc}

\begin{figure}[h]
	\begin{center}
		\includegraphics[width=0.45\textwidth]{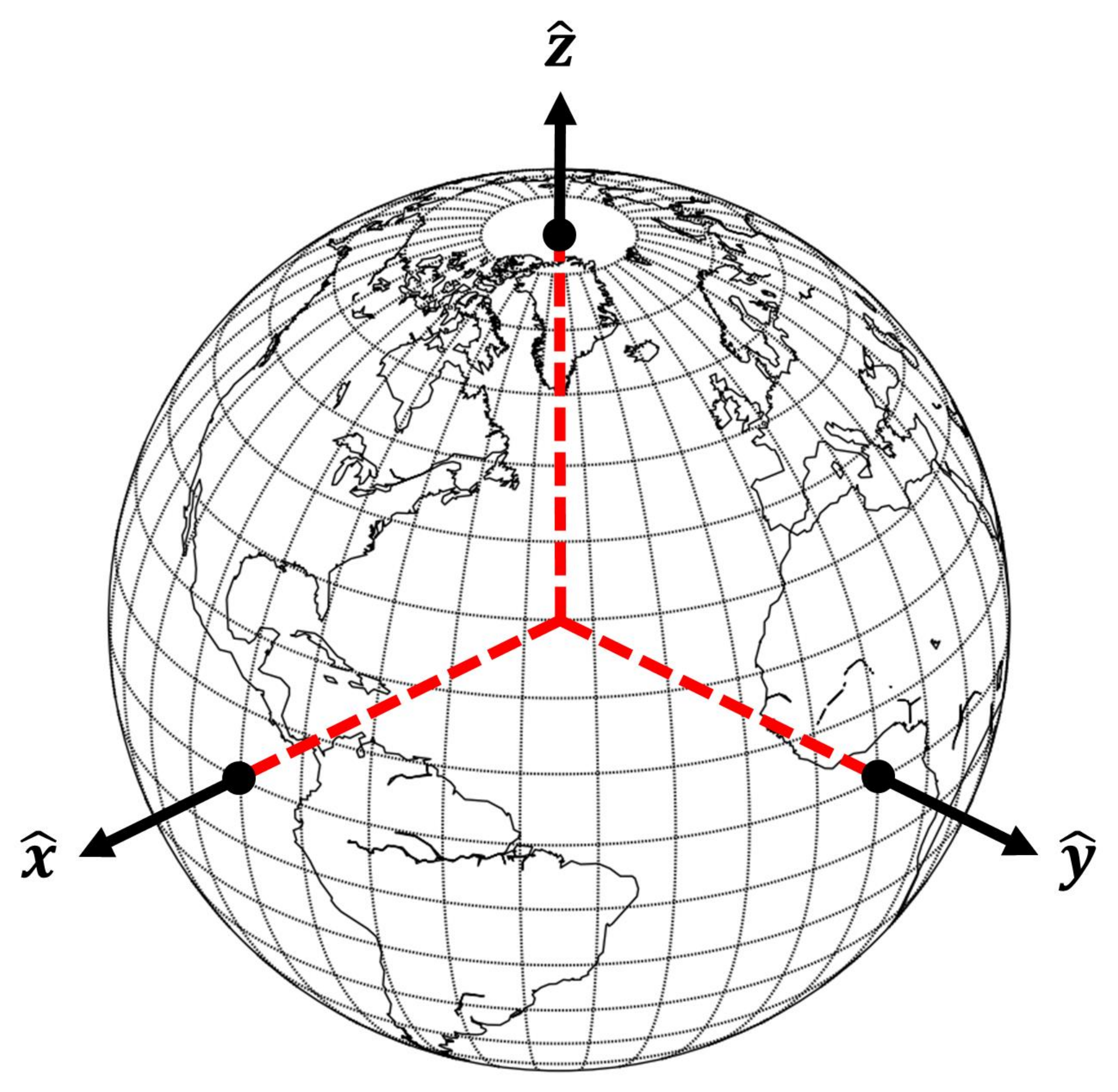}
		\caption{Definition of terrestrial axes used in \S \ref{subsect:atmodistro}.  The $+x$-axis passes through the 
		equator at the 90$^{\circ}$ W-meridian, the $+y$-axis passes through the equator at the 0$^{\circ}$ meridian, 
		and the $+z$-axis passes through the geographic North Pole.
		\label{fig:axesdefn}
		}
	\end{center}
\end{figure}

We define the Earth's terrestrial frame with the $+x$-axis passing through the equator at the 90$^{\circ}$ W-meridian, 
the $+y$-axis passing through the equator at the 0$^{\circ}$ meridian, and the $+z$-axis passing through the 
North Pole, as shown in Figure \ref{fig:axesdefn}).  We define spherical coordinates with $\theta$ as the 
polar angle from the $+z$-axis, $\phi$ as the azimuthal angle from the $+x$-axis, and $r$ as the 
radial distance from the center of the Earth:
\beq
\begin{pmatrix} x \\ y \\ z \end{pmatrix} = \begin{pmatrix} r \sin \theta  \cos⁡\phi \\ r \sin⁡\theta \sin⁡\phi \\ r \cos ⁡\theta \end{pmatrix} \, .
\eeq
We transform the terrestrial frame to Earth's rotating frame ($'$) by rotating about the $z$-axis with an 
angular speed of $\omega = 7.3 \times 10^{-5}$ rad s$^{-1}$ ($360^{\circ}/1 \ {\rm day}$), see Figure \ref{fig:xfrm01}:
\beq
\begin{pmatrix} x' \\ y' \\ z' \end{pmatrix} = \begin{pmatrix} \cos \omega t & -\sin \omega t & 0 \\ \sin \omega t & \cos \omega t & 0 \\ 0 & 0 & 1 \end{pmatrix} \begin{pmatrix} x \\ y \\ z \end{pmatrix} \, .
\label{eq:terrrot}
\eeq
Next we transform to the inclination frame ($''$) by rotating about the $x'$-axis by an angle $\alpha = 23.3^{\circ}$,
see Figure \ref{fig:xfrm12}:
\beq
\begin{pmatrix} x'' \\ y'' \\ z'' \end{pmatrix} = \begin{pmatrix} 1 & 0 & 0 \\ 0& \cos \alpha & \sin \alpha \\ 0 & -\sin \alpha & \cos \alpha \end{pmatrix} \begin{pmatrix} x' \\ y' \\ z' \end{pmatrix} \, .
\label{eq:rotincl}
\eeq
The next transformation is to the precessing/Ecliptic frame ($'''$) by rotating about the $z''$-axis with an angular speed
of $\chi = 7.7 \times 10^{-12}$ rad s$^{-1}$ ($360^{\circ}/26 \ {\rm kyr}$), see Figure \ref{fig:xfrm23}:
\beq
\begin{pmatrix} x''' \\ y''' \\ z''' \end{pmatrix} = \begin{pmatrix} \cos \chi t &\sin \chi t & 0 \\ -\sin \chi t & \cos \chi t & 0 \\ 0 & 0 & 1 \end{pmatrix} \begin{pmatrix} x'' \\ y'' \\ z'' \end{pmatrix} \, .
\label{eq:inclprec}
\eeq
Finally, we transform to the shock wave/interstellar frame ($''''$) by rotating about the $x'''$-axis by an angle $\eta$ 
to account for different directions of arrival, see Figure \ref{fig:xfrm34}:
\beq
\begin{pmatrix} x'''' \\ y'''' \\ z'''' \end{pmatrix} = \begin{pmatrix} 1 & 0 & 0 \\ 0 & \cos \eta & -\sin \eta \\ 0 & \sin \eta & \cos \eta \end{pmatrix} \begin{pmatrix} x''' \\ y''' \\ z''' \end{pmatrix} \, .
\label{eq:precshok}
\eeq
The arrival angle, $\eta$, is defined as the angle from the Ecliptic North Pole to the SN source.  In the interstellar frame, 
the particles travel along the $-\hat{z}''''$-direction, or:  $\vec{\mathbb{F}}(t)= -\mathbb{F}(t) \hat{z}''''$.  
We also define spherical coordinates in the interstellar frame so that:
\beq
\begin{pmatrix} x'''' \\ y'''' \\ z'''' \end{pmatrix} = \begin{pmatrix} r'''' \sin \theta'''' \cos \phi'''' \\ r'''' \sin \theta'''' \sin \phi'''' \\ r'''' \cos \theta'''' \end{pmatrix} \, .
\eeq
Combining the transformations we have:
\beq
\begin{pmatrix} x'''' \\ y'''' \\ z'''' \end{pmatrix} = \begin{pmatrix} 1 & 0 & 0 \\ 0 & \cos \eta & -\sin \eta \\ 0 & \sin \eta & \cos \eta \end{pmatrix} \begin{pmatrix} \cos \chi t & \sin \chi t & 0 \\ -\sin \chi t & \cos \chi t & 0 \\ 0 & 0 & 1 \end{pmatrix} \begin{pmatrix} 1 & 0 & 0 \\ 0 & \cos \alpha & \sin \alpha \\ 0 & -\sin \alpha & \cos \alpha \end{pmatrix} \begin{pmatrix} \cos \omega t & -\sin \omega t & 0 \\ \sin \omega t & \cos \omega t & 0 \\ 0 & 0 & 1 \end{pmatrix} \begin{pmatrix} r \sin \theta  \cos \phi \\ r \sin \theta  \sin \phi \\ r \cos \theta \end{pmatrix} \, .
\label{eq:transform}
\eeq

\begin{figure*}[t]
	\centering
	\subfigure[]
		{\includegraphics[width=0.22\textwidth]{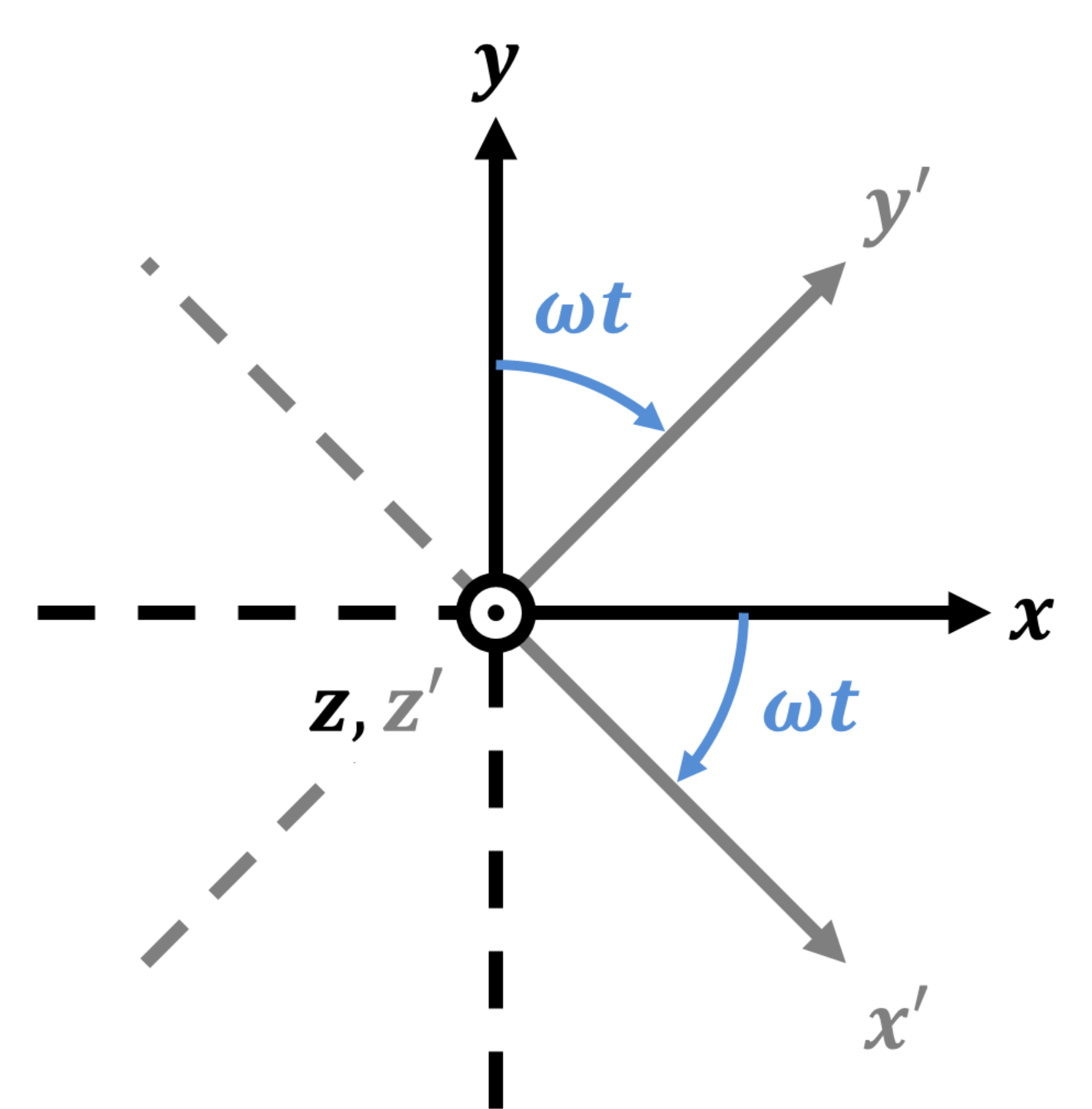}
		\label{fig:xfrm01}} ~
	\subfigure[]
		{\includegraphics[width=0.22\textwidth]{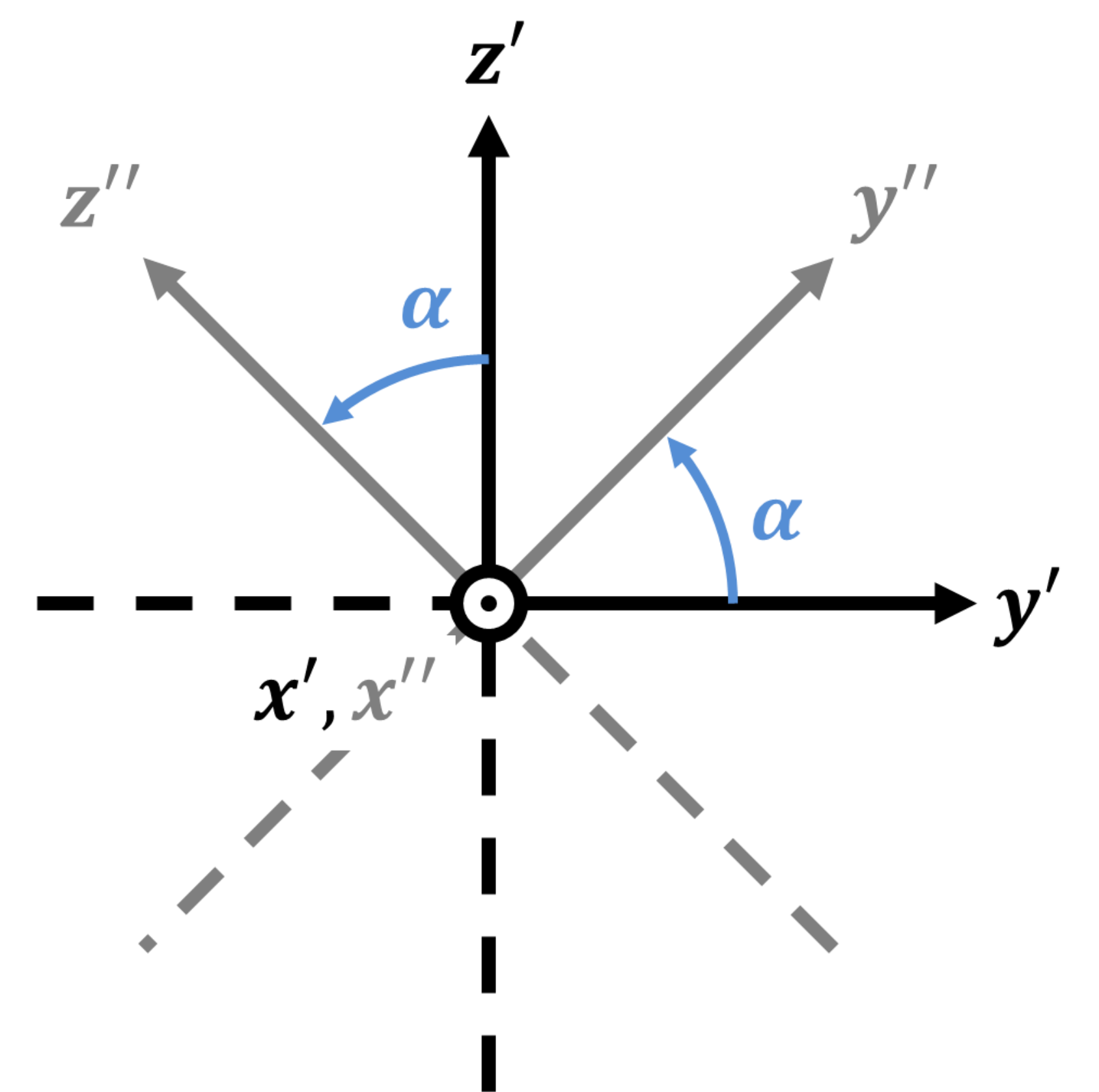}
		\label{fig:xfrm12}} ~
	\subfigure[]
		{\includegraphics[width=0.22\textwidth]{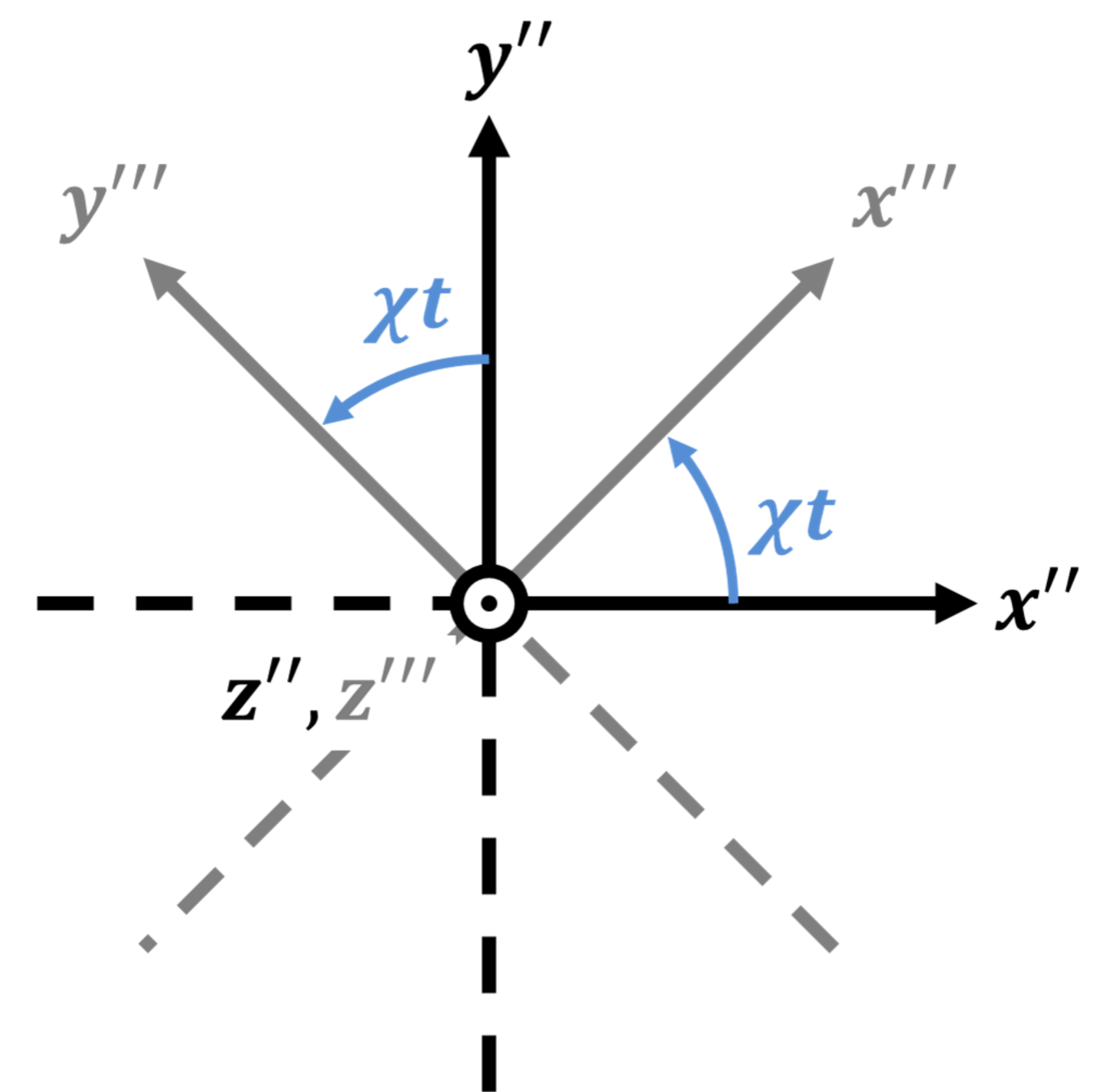}
		\label{fig:xfrm23}} ~
	\subfigure[]
		{\includegraphics[width=0.22\textwidth]{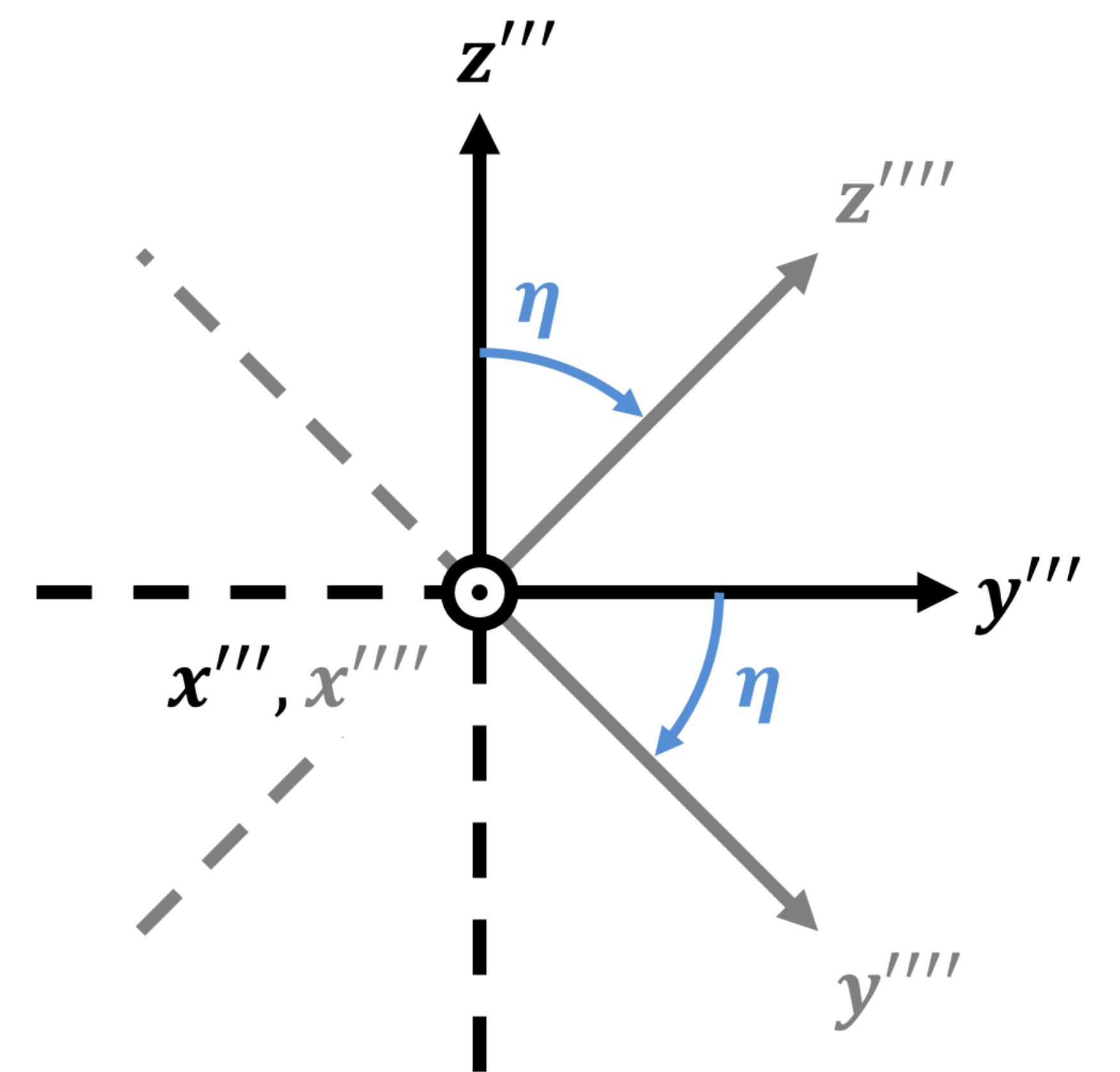}
		\label{fig:xfrm34}}
	\caption{Coordinate Transformations:  (a)  Terrestrial (unprimed) to Daily Rotation ($'$), (b)  Daily Rotation ($'$) to Inclination ($''$), (c)  Inclination ($''$) to Precessing/Ecliptic ($'''$), (d)  Precessing/Ecliptic ($'''$) to Shock Wave/Interstellar ($''''$).}
\end{figure*}

With regards to the coordinate differentials, all of the coordinate transformations are rotations, which means they are special affine transformations and therefore are area- and volume-preserving.  More specifically, examining the terrestrial-to-rotating frame transformation, the differential volumes are:
\beq
dV = d \vec{z} \cdot ( d \vec{x} \times d \vec{y} ) = dx ~dy ~dz \, {\rm , and \ } dV' = d \vec{z'} \cdot ( d \vec{x'} \times d \vec{y'} ) = dx' ~dy' ~dz' \, ,
\label{eq:difvolume}
\eeq
and the two sets of differentials are related according to Equation (\ref{eq:terrrot}):
\begin{align}
d \vec{x'} &= d \vec{x} \cos \omega t - d \vec{y} \sin \omega t \nonumber \\
d \vec{y'} &= d \vec{x} \sin \omega t + d \vec{y} \cos \omega t \nonumber \\
d \vec{z'} &= d \vec{z} \, .
\label{eq:difvec}
\end{align}
Combining Equations (\ref{eq:difvolume}) and (\ref{eq:difvec}), we get:
\begin{align}
dV' &= d \vec{z'} \cdot ( d \vec{x'} \times d \vec{y'} ) = d \vec{z} \cdot \left[ \left( d \vec{x} \cos \omega t - d \vec{y} \sin \omega t \right) \times \left( d \vec{x} \sin \omega t + d \vec{y} \cos \omega t \right) \right] \nonumber \\
    &= d \vec{z} \cdot \left[ \left( \cos^2 \omega t + \sin^2 \omega t \right) d \vec{x} \times d \vec{y} \right] = d \vec{z} \cdot \left[ \left(1 \right) d \vec{x} \times d \vec{y} \right] = d \vec{z} \cdot \left( d \vec{x} \times d \vec{y} \right) = dV \nonumber \\
\Rightarrow dx' ~dy' ~dz' &= dx ~dy ~dz \, .
\end{align}
Similar derivations can be done for the each of the other transformations, and we find:
\beq
\Rightarrow dx ~dy ~dz = dx'''' ~dy'''' ~dz'''' = r^2 \sin \theta ~d \theta ~d \phi ~dr = (r'''')^2 \sin \theta'''' ~d \theta'''' ~d \phi'''' ~dr'''' \, .
\eeq
Because there is no variation in radius,  $\sin \theta ~d \theta ~d \phi = \sin \theta'''' ~d \theta'''' ~d \phi'''' \Rightarrow 
d \Omega = d \Omega''''$, and $d \vec{\Omega}$ is directed away from Earth's center.  To calculate the fluence, ${\cal F}$, received by a sector of Earth, 
we integrate over the area of the sector:
\beq
dN = \vec{\mathbb{F}}(t) \cdot d \vec{A} ~dt \Rightarrow \frac{dN}{A} = \frac{\vec{\mathbb{F}}(t) \cdot d \vec{A} ~dt}{A}, d \vec{A} = r^{2} d \vec{\Omega}
\eeq
\beq
\Rightarrow d{\cal F} = \frac{\vec{\mathbb{F}}(t) \cdot r^{2} d \vec{\Omega} ~dt}{r^{2} \Omega} = \frac{\vec{\mathbb{F}}(t) \cdot d \vec{\Omega} ~dt}{\Omega}
\eeq
\beq
\Rightarrow {\cal F} = \frac{\iiint \vec{\mathbb{F}}(t) \cdot d \vec{\Omega} ~dt}{\Omega}, \mathbb{F}(t) = \mathbb{F}_{0} \left( 1 - \frac{t}{\Delta t_{\rm signal}} \right) \, .
\eeq
In the interstellar frame, $\theta''''$ and $\phi''''$ do not depend on time, $t$:
\beq
{\cal F} = \frac{\iiint \vec{\mathbb{F}}(t) \cdot d \vec{\Omega}'''' dt}{\Omega} = \frac{1}{\Omega} \iiint - \mathbb{F}(t) \cos \left(\pi - \theta'''' \right) d \Omega'''' dt = \frac{1}{\Omega} \iiint \mathbb{F}(t) \cos \theta'''' d \Omega'''' dt
\eeq
\beq
\Rightarrow {\cal F} = \frac{1}{\Omega} \int_{t_{\rm ini}}^{t_{\rm fin}} \mathbb{F}(t) dt \iint_{S} \cos \theta'''' \sin \theta'''' d \theta'''' d \phi''''
\label{eq:fluencesector}
\eeq
The first integral is straightforward: 
\beq
\int_{t_{\rm ini}}^{t_{\rm fin}} \mathbb{F}(t) dt = \mathbb{F}_{0} \left[ t - \frac{t^{2}}{2 \Delta t_{\rm signal}} \right] _{t_{\rm ini}}^{t_{\rm fin}} = \mathbb{F}_{0} \left( \frac{t_{\rm ini}^{2} - t_{\rm fin}^{2}}{2 \Delta t_{\rm signal}} + t_{\rm fin} - t_{\rm ini} \right) \, ,
\label{eq:fluencemag}
\eeq
and the second integral is the projected area of a spherical sector onto the $x'''' - y''''$-plane, see Figure \ref{fig:sector}.  

\begin{figure}[t]
	\begin{center}
		\includegraphics[width=\textwidth]{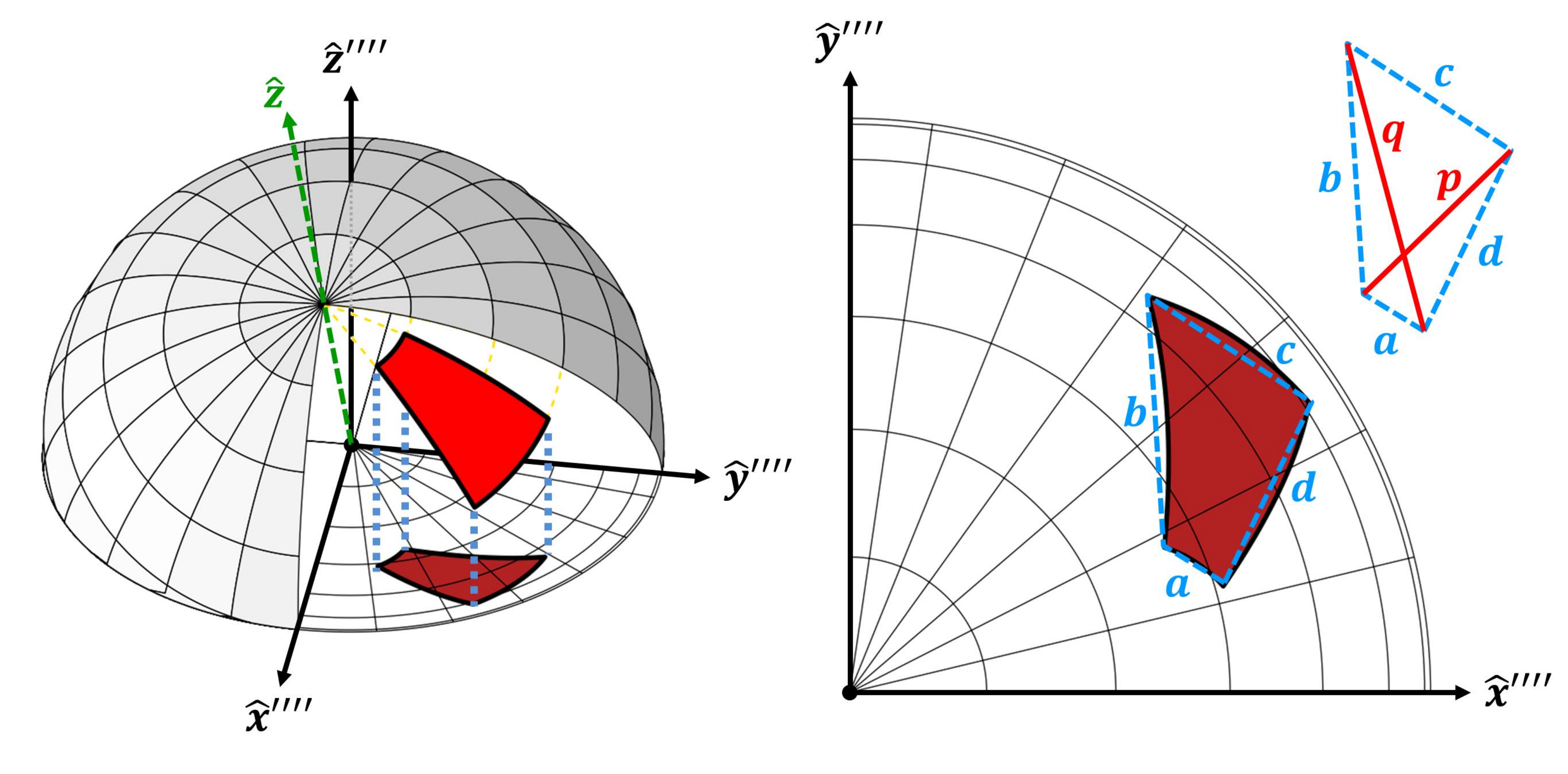}
		\caption{Sector area approximation.  The area of the projection of the sector onto the $x'''' - y''''$-plane is approximated using 
		Bretschneider's Formula for 4-sided sectors and Heron's Formula for 3-sided sectors.
		\label{fig:sector}
		}
	\end{center}
\end{figure}

Because the SN dust particles are traveling in the $-\hat{z}''''$-direction by construction, the surface integral in 
Equation (\ref{eq:fluencesector}) represents the area of the sector projected onto the $x'''' - y''''$-plane (see Figure \ref{fig:sector}).  
While it is fairly straightforward to transform the sector vertices from the terrestrial to the interstellar frame 
(e.g., $\theta_{u} \rightarrow \theta_{u}''''$, etc.) the path from each vertex is not, requiring a dependence on 
$\phi''''$ in the limits of integration for $\theta''''$ (or vice versa):
\beq
\iint_{S} \cos \theta'''' \sin \theta'''' d \theta'''' d \phi'''' = \int_{\phi_{l}''''}^{\phi_{u}''''} \int_{g(\phi'''')}^{f(\phi'''')} \cos \theta'''' \sin \theta'''' d \theta'''' d \phi'''' \, ,
\eeq
where $f$ and $g$ are the transformed paths between vertices.  In order to simplify our calculations, 
rather than derive the transformation functions, we approximate the area of the projection (and the integral) as a general quadrilateral (or triangle in the case where $\theta_{l} = 0^{\circ}$ or $\theta_{u} = 180^{\circ}$).  
In other words:
\beq
\iint_{S} \cos \theta'''' \sin \theta'''' d \theta'''' d \phi'''' \approx 
\begin{cases} 
\sqrt{s (s - a) (s - b) (s - c)} & \quad \text{if } \theta_{l} = 0^{\circ} \text{ or } \theta_{u} = 180^{\circ} \, , \\ \\
\frac{1}{4} \sqrt{4 ~p^{2} ~q^{2} - \left( b^{2} + d^{2} - a^{2} - c^{2} \right) ^{2}} & \quad \text{otherwise} \, , \\
\end{cases}
\label{eq:fluencearea}
\eeq
where $a$, $b$, $c$ and $d$ are the lengths of each side, $p$ and $q$ are the lengths of the diagonals of the quadrilateral, and $s$ is the semi-perimeter of the triangle ($s \equiv (a + b + c)/2$).

For a given latitude and longitude $(\theta, \phi)$ and gridsize $(\Delta \theta \times \Delta \phi)$, these correspond to grid boundaries at:  $\theta_{l} = \theta - \Delta \theta/2$, $\theta_{u} = \theta + \Delta \theta/2$, $\phi_{l} = \phi - \Delta \phi/2$, and $\phi_{u} = \phi + \Delta \phi/2$.  Using Equation (\ref{eq:transform}) and setting $r = 1 \ \RE$, we transform the grid vertices to their associated $(x'''', y'''', z'''')$ coordinates.  The distances between the $(x'''', y'''', 0)$ coordinates are used to find the associated $a$, $b$, $c$, $d$, $p$, and $q$ values.  
For our particular approach, the vertices correspond to:
\begin{align}
a ~& \widehat{=}~ \overline{\left( \phi_{l}, \theta_{l} \right) \left( \phi_{u}, \theta_{l} \right)}
&b ~& \widehat{=}~ \overline{\left( \phi_{u}, \theta_{l} \right) \left( \phi_{u}, \theta_{u} \right)}
&c ~& \widehat{=}~ \overline{\left( \phi_{u}, \theta_{u} \right) \left( \phi_{l}, \theta_{u} \right)} \nonumber \\
d ~& \widehat{=}~ \overline{\left( \phi_{l}, \theta_{u} \right) \left( \phi_{l}, \theta_{l} \right)}
&p ~& \widehat{=}~ \overline{\left( \phi_{u}, \theta_{l} \right) \left( \phi_{l}, \theta_{u} \right)}
&q ~& \widehat{=}~ \overline{\left( \phi_{l}, \theta_{l} \right) \left( \phi_{u}, \theta_{u} \right)} \, .
\end{align}
If any of the $z''''$ values for the vertices are negative, corresponding to the sector being on the 
opposite side of the Earth from the arriving flux, the area of the sector is zero.  This increases the error in 
our approximation along the edges, but our time intervals are such that the errors are consistent across the 
entire surface, and we are interested in the relative values across the globe.

Finally, to calculate the fluences for each of the sectors (for $\Delta \theta = \Delta \phi = 10^{\circ}$ there are 648 sectors), 
we calculate the fluence received at each time step on each sector then sum over all time steps for each sector (presumably the 
time steps cover the entire duration of the SN signal, although after one precessional cycle the final pattern changes very little).  
We use Equation (\ref{eq:fluencemag}) to find the fluence incident to the sector during that time step (given by Equations (\ref{eq:timeprec}) 
and (\ref{eq:timeday})), and we use Equation (\ref{eq:fluencearea}) to find the cross-sectional area facing the incident SN dust flux.  The product 
of these two values gives the sector fluence at each time step.  Once the sector fluence has been summed over all time steps (the result of 
Equation (\ref{eq:fluencesector})), the value of $\psi$ for each sector is found by scaling the total sector fluence by the total area-weighted average 
fluence of the entire sphere.  For our chosen grid size of $\Delta \theta = \Delta \phi = 10^{\circ}$, this approximation is accurate to $\lesssim 1\%$, and 
for our results in Figures \ref{fig:distro} and \ref{fig:moondep}, this approximation demonstrated convergence to the precision given.

\section{B.  Heliosphere IMF Model}
\label{app:imfmodel}

We use the model outlined in \citet{park58} and \citet{gust94}.  Using a right-handed, spherical coordinate system with the Sun at the origin, we define $\phi$ as the azimuthal angle along the Sun's equator and $\theta$ as the angle from the Sun's rotational axis.  Because of the Sun's rotation and a magnetic field frozen-in the radially expanding solar wind, the components of the IMF take the form:
\begin{align}
B_{r} &= B_{r, 0} \left( \frac{r_0}{r} \right)^{2} {\rm sgn}\left( \pi/2 - \theta \right) \, , \\
B_{\theta} &= 0 \, , \\
B_{\phi} &= B_{\phi, 0} \left( \frac{r_0}{r} \right) \sin{\theta} ~{\rm sgn}\left( \pi/2 - \theta \right) \, ,
\end{align}
where $B_{r, 0}$ and $B_{\phi, 0}$ are the magnetic field components at $r_0$ and ${\rm sgn}\left( \pi/2 - \theta \right)$ accounts for the different polarities in the northern and southern solar hemispheres.  From \citet{gust94}, for $r_0 = 1$ AU, $B_{r, 0} \approx B_{\phi, 0} \approx 30 \ \mu$G.

\section{C.  Earth's Magnetosphere Model}
\label{app:msmodel}

The \citet{kats87} Magnetosphere model defines its right-handed axes with the origin at the Earth, 
the $\hat{X}$-axis towards the Sun, the $\hat{Z}$-axis through the geographic North, and takes the form:
\beq
\vec{B} = \vec{B}_{\rm dipole} + \vec{B}_{\rm tail} \, ,
\eeq
with:
\beq
\vec{B}_{\rm dipole} = - \frac{3MZX}{R^{5}} \hat{X} - \frac{3MZY}{R^{5}} \hat{Y} + \left( \frac{M}{R_{3}}-\frac{3MZ^{2}}{R^{5}} \right) \hat{Z} \, ,
\eeq
\beq
\vec{B}_{\rm tail} = \left( B_{X0} \tanh \left( Z/H \right) \right) \hat{X} - B_{Z0} \hat{Z} \, ,
\eeq
and
\beq
R^{2} = X^{2} + Y^{2} + Z^{2} \, ,
\eeq
where $M$ is the magnetic dipole strength based on the surface value, and $B_{Z0}$ is a uniform magnetic field 
normal to the dipole's equatorial plane.

\end{document}